# SPATIO-TEMPORAL CHARACTERIZATION OF CAUSAL ELECTROPHYSIOLOGICAL ACTIVITY STIMULATED BY SINGLE PULSE FOCUSED ULTRASOUND: AN EX VIVO STUDY ON HIPPOCAMPAL BRAIN SLICES

Ivan M. Suarez-Castellanos[1], Elena Dossi[2], Jeremy Vion-Bailly[1], Lea Salette[2], Jean-Yves Chapelon[1], Alexandre Carpentier[4,5], Gilles Huberfeld[2,3], William Apoutou N'Djin[1]

[1] LabTAU, INSERM, Centre Léon Bérard, Université Lyon 1, Univ Lyon, F-69003, Lyon, France

[2] Neuroglial Interactions in Cerebral Physiopathology, Center for Interdisciplinary Research in Biology, Collège de France, CNR UMR 7241, INSERM U1050, Paris, France

[3] AP-HP, Neurophysiology department, Sorbonne University, Pitié-Salpêtrière Hospital, Paris, France

[4] AP-HP, Neurosurgery department, Pitié-Salpêtrière Hospital, Paris, France

[5] Sorbonne University, GRC23, Interface Neuro Machine team, Paris, France

## Abstract

Objective: The brain operates via generation, transmission and integration of neuronal signals and most neurological disorders are related to perturbation of these processes. Neurostimulation by Focused Ultrasound (FUS) is a promising technology with potential to rival other clinically-used techniques for the investigation of brain function and treatment of numerous neurological diseases. The purpose of this study was to characterize spatial and temporal aspects of causal electrophysiological signals directly stimulated by short, single pulses of focused ultrasound (FUS) on *ex vivo* mouse hippocampal brain slices.

Approach: MicroElectrode Arrays (MEA) are used to study the spatio-temporal dynamics of extracellular neuronal activities both at the single neuron and neural networks scales. Hence, MEAs provide an excellent platform for characterization of electrical activity generated, modulated and transmitted in response to FUS exposure. In this study, a novel mixed FUS/MEA platform was designed for the spatio-temporal description of the causal responses generated by single 1.78 MHz FUS pulses in *ex vivo* mouse hippocampal brain slices.

Main results: Our results show that FUS pulses can generate local field potentials (LFPs), sustained by synchronized neuronal post-synaptic potentials, and reproducing network activities. LFPs induced by FUS stimulation were found to be repeatable to consecutive FUS pulses though exhibiting a wide range of amplitudes (50 – 600 µV), durations (20 - 200 ms), and response delays (10 – 60 ms). Moreover, LFPs were spread across the hippocampal slice following single FUS pulses thus demonstrating that FUS may be capable of stimulating different neural structures within the hippocampus.

Significance: Current knowledge on neurostimulation by ultrasound describes neuronal activity generated by trains of repetitive ultrasound pulses. This novel study details the causal neural responses produced by single-pulse FUS neurostimulation while illustrating the distribution and propagation properties of this neural activity along major neural pathways of the hippocampus.

**Keywords:** Focused Ultrasound (FUS), Single Pulse, Neurostimulation, MicroElectrode Arrays (MEA), Ex-Vivo Mouse Hippocampus, Local Field Potential (LFP)

# Introduction

Neurostimulation/neuromodulation refers to the group of technologies aimed at purposefully stimulating (i.e. producing de novo) or modulating (i.e. modifying a preexisting) neuronal activity for exploration of neuronal activities and treatment of various pathologies of neuronal or glial nature such as Parkinson's disease, epilepsy, depression disorder, among others. Focused Ultrasound (FUS) neurostimulation/neuromodulation has become, over the years, the subject of much research and investigation because of its capability to be applied non- or minimally-invasively and its inherent ability to target deep volumes in the cubic-millimeter range through geometric or electronic focusing of the ultrasound beam (Denworth, 2018; Yoo, 2018). Over the past several years, various ultrasound neurostimulation studies on animal models have reported on movement of limbs, whiskers, as well as eye movements and pupil dilation when selectively stimulating relevant motor regions of the animal's brain with ultrasound (King et al., 2012; Yoo et al., 2012; Younan et al., 2013; Kamimura et al., 2015). Despite the promise of this technology and significant recent advances, the development of this therapeutic modality into a clinical setting remains partially limited by the lack of understanding of the cellular and biomolecular mechanisms that mediate the electrophysiological responses of neurons to ultrasound energy. Our group recently presented an *in vivo* invertebrate nervous model of earthworm for study of electrophysiological mechanisms involved in FUS neurostimulation (Vion-Bailly et al., 2019). However, despite the simplicity and practicality of this model, it does not provide information on the spatial distribution of FUS-stimulated electrophysiological signals within neuronal networks and does not allow specific analysis of generated vs propagated signal. Other previous studies have used *in vitro* and *ex vivo* setups commonly-used in neurosciences such as whole-cell patch clamp and fluorescence imaging for characterization of the electrophysiological properties of neurostimulation (Tyler et al., 2008; Kubanek et al., 2016; Prieto et al., 2017). However, while the patch-clamp technique does not allow for studies of long-range connectivity in neural networks, fluorescence microscopy often lacks temporal resolution for proper characterization of neural events in the microsecond to millisecond range.

The hippocampal formation is based on a neuronal network that has been extensively described in literature (Andersen et al., 2006). Indeed, the effects of electrical stimulation, recording and analysis of electrophysiological activity in *ex vivo* hippocampal brain slices are well-known techniques that have been widely used for studies of synaptic transmission across major neural pathways of the hippocampus (Kopanitsa et al., 2006; Chever et al., 2016). Extracellular recordings of hippocampal electrical activity include local field potentials (LFPs) which are transient electrical signals generated by the summation and synchronization of the electrical synaptic activity of neurons. Stimulation of Schaffer's collateral, the axons of CA3 pyramidal cells connecting CA1 neurons, elicits a fiber volley (FV), which is an indication of the presynaptic action potential arriving at the recording site and a delayed field excitatory postsynaptic potential (fEPSP). Extracellularly recorded fEPSPs are associated with neurotransmitter-induced changes in the $Na^+$ and $K^+$ conductance of the postsynaptic neuron, thus creating ionic currents across the cell membrane that can lead to its depolarization in dendrites, (Andersen et al., 2006) and generates an electrical dipole. Previous studies have combined ultrasound systems with hippocampal slices for exploration of the modulating capabilities of ultrasound energy on electrically stimulated activity in CA1 and the Dentate Gyrus (Rinaldi et al., 1991; Bachtold et al., 1998). These studies demonstrated that ultrasound therapy is capable of significantly reducing the amplitude of the FV and cell population potentials, while enhancing dendritic potentials. In this study, however, we sought to study the capabilities of ultrasound energy to create electrical responses in these neural pathways. Studying and understanding the effects of ultrasound neurostimulation on triggering LFPs and their transmission along major neural pathways are mandatory for the development of potential clinical applications aiming at reinforcing neural pathways, secondary to damages or at quantifying functional neural connectivity.

Microelectrode Arrays (MEAs) are devices that contain a matrix of tens to hundreds of spatially-distributed microelectrodes that allow for the simultaneous delivery and real-time recording of extracellular neural signals ranging from isolated cultured cells to acute brain slices. MEA systems provide an excellent platform for the study of spatio-temporal dynamics of neural networks (Obien et al., 2015). This technology has been widely used for studies of brain function and neural signal transfer and processing (Menendez de la Prida and Huberfeld, 2019), (Streit et al., 2010), (Dossi et al., 2014). Several studies have used MEA systems to record actions potentials from multiple neurons (Multi Unit Activity) as well as synchronized synaptic potentials (Local Field Potentials) (Obien et al., 2015). MEA systems are particularly well-suited for extracellular recordings of hippocampal activity as neurons in this region are arranged in tightly delineated layers, and therefore, signals arising from evoked responses from large populations of neurons don't cancel out but rather add up to a signal that is easily detected by the electrodes in the chip (Cooke and Bliss, 2006). As a result, this platform can provide a useful and invaluable tool for description and characterization spatio-temporal patterns of neural activity created by exposure to FUS in *ex vivo* mouse hippocampal brain slices. Moreover, the array of electrodes lying under the tissue slice, it allows full access to the FUS device on the top side.

In this study, we demonstrated the feasibility of using a MEA platform integrated to a FUS system for the investigation of hippocampal neural signals generated by single-pulse FUS stimulation. The temporal and spatial characteristics of the responses that can be generated from FUS-exposed hippocampal brain slices and recorded with MEA systems were analyzed and described.

## Methods and Materials

### Ethic statements

All experiments were carried out on *ex-vivo* brain tissues. To minimize the number of sacrificed animals, *ex vivo* experiments were performed on 6 sacrificed mice from an ongoing *in vivo* protocol carried out in strict accordance with the legal conditions of the French National Ethics Committee for Reflection on Animal Experimentation (CNREEA). According to the last directive (decree 2013-118 and decrees from 1/02/2013), the *in-vivo* protocol (ref. APAFIS#912-20 150624150 15717 v2) was approved by the French Ministry of Higher Education and Research, and by the Paris local Ethics Committee on Animal Experiments (CNREEA Code: 2015-18 #1119). The presented *ex-vivo* study was then performed on an adult mouse model (C57Bl6) in the College de France (CIRB - CNRS UMR 7241 / INSERM U1050).

### Slice Preparation

Eight hippocampal slices from 5 adult mice were used in this study and were prepared as described by Dossi et al. and placed in conditions interfacing FUS treatment and recording using a MultiElectrodeArray (MEA) system (Dossi et al., 2014). Briefly, the mice were first anesthetized. The mice were then sacrificed by decapitation and their brains extracted for subsequent slicing using a vibratome (Leica VT1200, Leica Microsystems), according to standard procedures described in previous studies (Chever et al., 2016). Four hundred micrometer transverse slices were obtained, sagittally halved with a scalpel and immediately transferred to an interface chamber (BSC 2, Scientific System Design GmbH, Hofheim, Germany) for slice incubation (**Figure 1a-c**). The interface chamber consists of a temperature and oxygen-controlled chamber allowing perfusion of the brain slices with preoxygenated aCSF containing 124 mM NaCl, 3 mM KCl, 26 mM $NaHCO_3$, 10 mM D-Glucose, 1.6 mM $CaCl_2$, 1.3 mM $MgCl_2$, while maintaining a constant temperature of 35°C and high oxygen tension by oxygenation with carbogen (95% $O_2$ and 5% $CO_2$). Immediately before FUS treatment, a slice was transferred to the MEA chip (60 contacts, 200 µm intercontact spacing) for subsequent signal recording using an MEA system (MEA2100, MultiChannel Systems, Reutlingen, Germany). The chip was constantly perfused with preoxygenated aCSF preheated at 35°C (**Figure 1d**). The slice was placed so that the electrode array was located in the vicinity of known hippocampal structures such as CA1, CA3 and the dentate gyrus (**Figure 1e-f**). A weight with a grid transparent to both light and ultrasound was placed over the brain slice to ensure proper electrical contact between the slice and the electrodes, and to immobilize the slice during ultrasound experiments. Depending on the site of ultrasound stimulation in the slice, we aimed to differentiate activations of neurons soma (pyramidal layer of CA3, CA1), neuron dendrites (stratum radiatum of CA3), bundle of axons (perforant path or Schaffer collaterals) and propagation of activities (stimulation in CA3 / Schaffer collaterals – downstream recording in CA1).

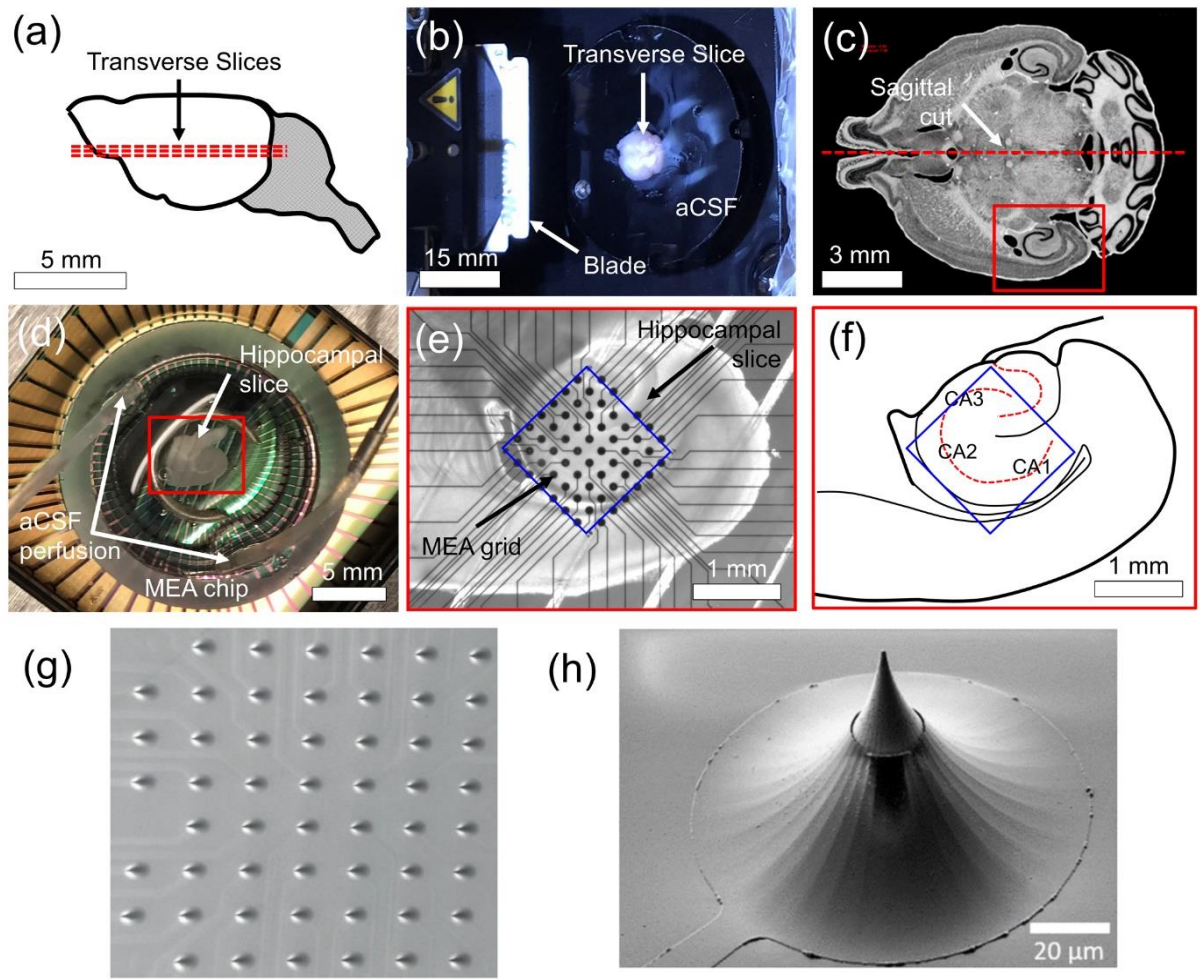

**Figure 1** - *Brain slices and MEA recording preparations. (a) Schematic of the brain (side view) to be sliced transversally every 400 µm. (b) Brain slicing using a vibratome. (c) Transverse cortical slice (Mouse Brain Atlas: C57BL/6J Horizontal, Atlas 232; http://www.mbl.org/atlas232/atlas232_frame.html). (d) Macroscopic view of a half transverse brain slice placed inside an aCSF-perfused 3D MEA chip. (e) Microscopic view of the MEA grid and anatomical structure of the hippocampal slice. (f) Schematic of the MEA grid position with respect to the CA1, CA2 and CA3 regions. (g) Closeup on the 3D electrode array of the MEA chip and (h) a single 3D electrode (https://www.multichannelsystems.com/sites/multichannelsystems.com/files/documents/manuals/MEA_Manual.pdf).*

## MEA System

The MEA chip includes an 8-by-8 grid of 60 TiN 3D electrodes (the four corners of the grid do not contain electrodes) with a 12 µm diameter and 200 µm interelectrode spreading (**Figure 1d-g**). The 3D electrodes in this type of MEA chip (base diameter: 100 µm, tip diameter: 12 µm) can penetrate 50 µm deep into the brain slice (**Figure 1h**), thus improving spatial recording and contributing to immobilize the slice, thereby reducing FUS associated mechanical noise. The raw electrical activity detected by the electrodes in the MEA chip was recorded using software developed by MultiChannel Systems (MC_Rack V 4.6.2, MultiChannel Systems GmbH, Reutlingen, Germany) at a sampling rate of 50 kHz and filtered with a low-pass filter with cutoff frequency of 10 kHz and a 0.1 Hz high-pass filter. The noise level, as estimated through calculation of the RMS noise, was calculated to be 8.97 $\mu V_{RMS}$ which implies peak-to-peak noise to be approximately 59.2 µV. The excitability of the brain slice was first checked by electrically stimulating (5 $V_{peak-peak}$ biphasic pulse: 100 µs negative and 100 µs positive, increased intensity up to a significant response) Schaffer collaterals in CA3 and observing resulting responses in stratum pyramidale and radiatum in CA1 (Chever et al., 2016).

## FUS Setup

A FUS stimulation system was integrated to the MEA system to form a mixed FUS-MEA platform allowing us to stimulate the hippocampal slices while simultaneously recording their electrical activity in real-time. The ultrasound stimulation system consisted of a custom-made single-element focused transducer (PZ26, Meggitt A/S, Kvistgaard, Denmark), with resonant frequency of 1.78 MHz, thickness of 1.25 mm, diameter of 15 mm and radius of curvature of 15 mm (**Figure 2**). We chose a PZT ceramic resonating at a frequency higher than those usually utilized in the field of FUS neurostimulation (< 1 MHz) because of its shorter wavelength providing higher spatial selectivity and thus finer focusing (Kamimura et al., 2015). Such features could be of interest for diverse FUS-brain clinical applications that integrate technology to bypass the attenuating skull (Carpentier et al., 2016; Prada et al., 2020). The transducer was driven by an electrical waveform generated using a Tektronix TDS3014 function generator (Tektronix Inc., Beaverton, Oregon, USA) and amplified using a 50 dB amplifier (Kalmus 150 RF, Amplifier Research Modular RF, Bothell, WA, USA). Two challenges were immediately addressed: acoustic coupling between the transducer and the extracellular aCSF in the MEA chip, and electronic coupling of all the electrical equipment included in the setup. Acoustic coupling was achieved by using a 3D-printed, acrylonitrile butadiene styrene (ABS) cone filled with 0.2% agarose gel to guide the generated ultrasound wave to the perfused fluid inside the MEA chip. The attenuation coefficient for such a low agarose concentration is below 0.04 dB/cm as reported by previous studies (Culjat et al., 2010; Yang et al., 2017). Furthermore, since the density, speed of sound and thus acoustic impedance of low-concentration agar-based gels are very close to those of water (1016 kg/m$^3$, 1498 m/s, 1.52 MRayls, respectively)(Burlew et al., 1980), ultrasound reflections at the gel-aCSF boundary were estimated to be negligible (<1%). The cone was designed to match the focusing path of the ultrasound emitted by the FUS transducer. The top aperture of the cone (2 mm in diameter) was designed to be located 2 mm away from the focus of the transducer. Electronic coupling was ensured by connecting the masses of the FUS and MEA systems. Furthermore, a polycrilamide envelope was placed around the cone to physically isolate the aCSF and MEA electrodes from the FUS transducer and the coupling gel. The ultrasound transducer and its waveguide were carefully positioned inside the MEA chip and directed towards the hippocampus of the brain slice inside the chip using a micropositioner. Targeting of the hippocampal area in direct contact with the electrode matrix was done using an inverted camera included with the MEA system. The rim of the cone edge was painted white to enhance its visibility in the real-time video provided by the MEA camera. This facilitated the positioning of the focal spot on the hippocampal slice since the focal spot was located at the virtual tip of the guiding cone (**Figure 2a and d**).

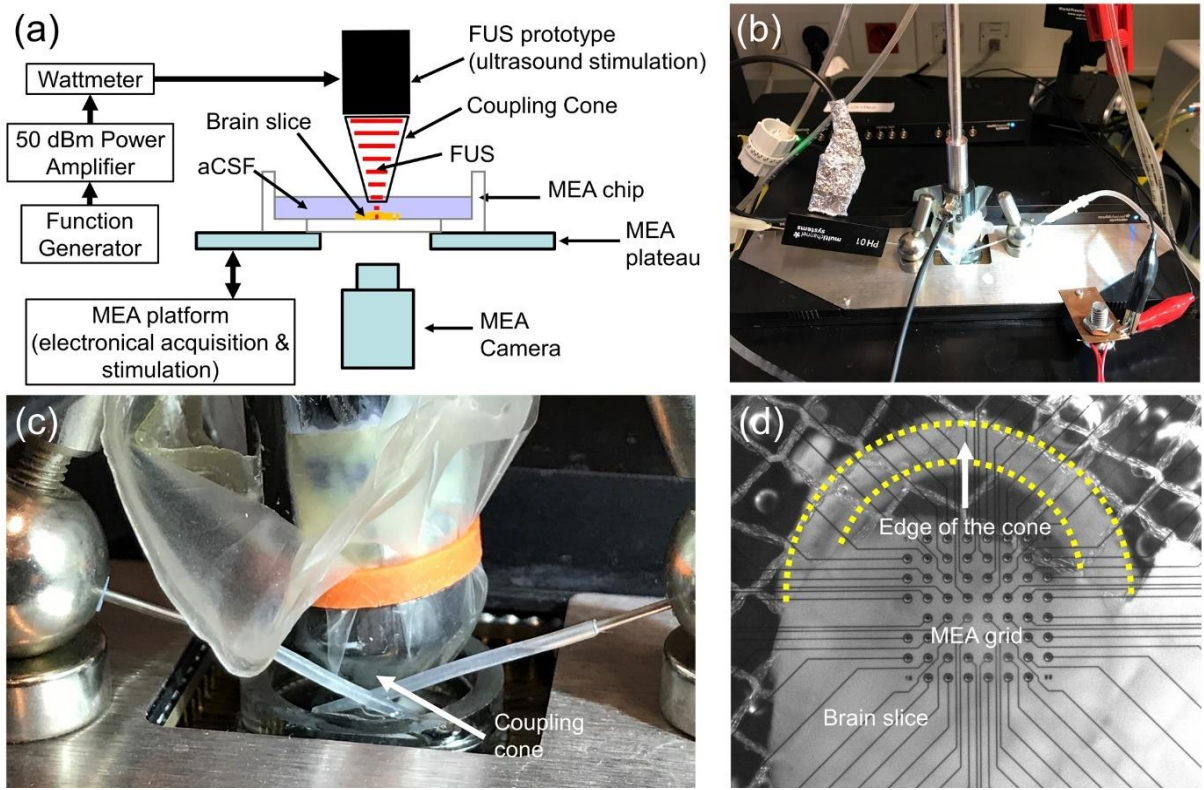

**Figure 2** - *Mixed MEA-FUS Stimulation platform.* ***(a)*** *Schematic of the experimental set-up.* ***(b)*** *Picture of the experimental set-up.* ***(c)*** *Zoom in on the FUS prototype in acoustic coupling with the MEA chip.* ***(d)*** *Bottom view (MEA camera) of the MEA grid (first plan) + the brain slice (second plan) + the coupling cone (background).*

## FUS Transducer Characterization

In the manuscript, the XYZ orthonormal referential was oriented such that the XY plane was perpendicular to the direction of propagation of ultrasound along the Z axis. As such, ultrasound was administrated in normal incidence to the brain slices located in the XY plane. The time-average acoustic power, $P_{ta}$, generated by the FUS transducer was estimated by measuring the radiation force exerted by the acoustic wave on an absorber mounted to the weighing plate of a high precision balance (RFB-2000 and standard absorbing target, Onda Corp., Sunnyvale, CA, USA) (Davidson, 1991). Using the measured $P_{ta}$, we calculated the pulse-average power which was used to estimate the pressures generated for stimulation of the brain slice. The testing setup for measuring acoustic power consisted of immersing the emitting surface of the FUS transducer in degassed water and positioning it in parallel to the acoustic absorber. Various 1.78 MHz sinewaves of different input amplitudes were applied on the transducer to determine the acoustic radiation force that could be generated by the FUS transducer.

Pressure field measurements were performed by placing the FUS transducer in a tank filled with degassed water. A calibrated hydrophone (HGL-0200, Onda Corp., Sunnyvale, CA, USA) was attached to a 3D micropositioning system, submerged in the water of the tank and oriented to have its sensitive electrode surface face the surface of the FUS transducer. The scan acquisitions were taken at 0.1 mm intervals in the length-width dimensions (perpendicular to US propagation) and 0.2 mm steps in the depth dimension (US propagation direction) (Bawiec et al., 2018). The size of the FUS focal spot was determined as the -3 dB region relative to the maximum pressure measured (**Figure 3c**). In normal incidence of the beam, the focal spot was circular and 0.8 mm in diameter which provided an expected area of stimulation at the focus. The pressure along the acoustic propagation axis was found to vary by no more than 2 dB over a distance of ± 2 mm away from the location of maximum deposited pressure (**Figure 3d**). This is relevant to our experiment since we expect the brain slice to be located somewhere along the distance separating the location of maximum pressure and the top edge of the coupling cone (approximately 2 mm).

FUS pressures applied throughout the study and reported spatial-average RMS acoustic pressures at the -3 dB focal spot ($p_{sar}$, Preston, 1991), unless otherwise stated, ranged from 2.5 to 8 MPa, the pulse-repetition frequency (PRF) varied from 0.2 to 1 Hz, and the pulse duration varied from 160 to 200 µs (**Figure 3b**). Only in the study aimed at evaluating the effects of TTX on the FUS-induced responses were the pulse durations and PRF set at 3.5 ms and 4 Hz, respectively. When activating FUS while the transducer was immersed in aCSF, electromagnetic interferences (EMI) were always visible on all electrodes of the highly sensitive MEA (amplifying platform). However, neural responses were only created when the FUS focal spot was placed at the front of the targeted brain structure (see study in **Figure 8d**). The EMI formed a "FUS artifact" on the recording data, which was used as a time reference for analyses. Application of short-duration pulses prevented interference between recorded FUS-induced artifacts and the elicited responses. Moreover, utilization of pulse durations in the microsecond range allows for direct comparison with other neurostimulation techniques that use similar pulse durations (Miniussi et al., 2012). FUS stimulation was centered on the same area as electrical stimulation, although the stimulated area by the current device was wider. The focal spot was preferentially positioned to target CA3 while particular attention was placed on recording responses in electrodes located within CA1 (**Figure 3c**).

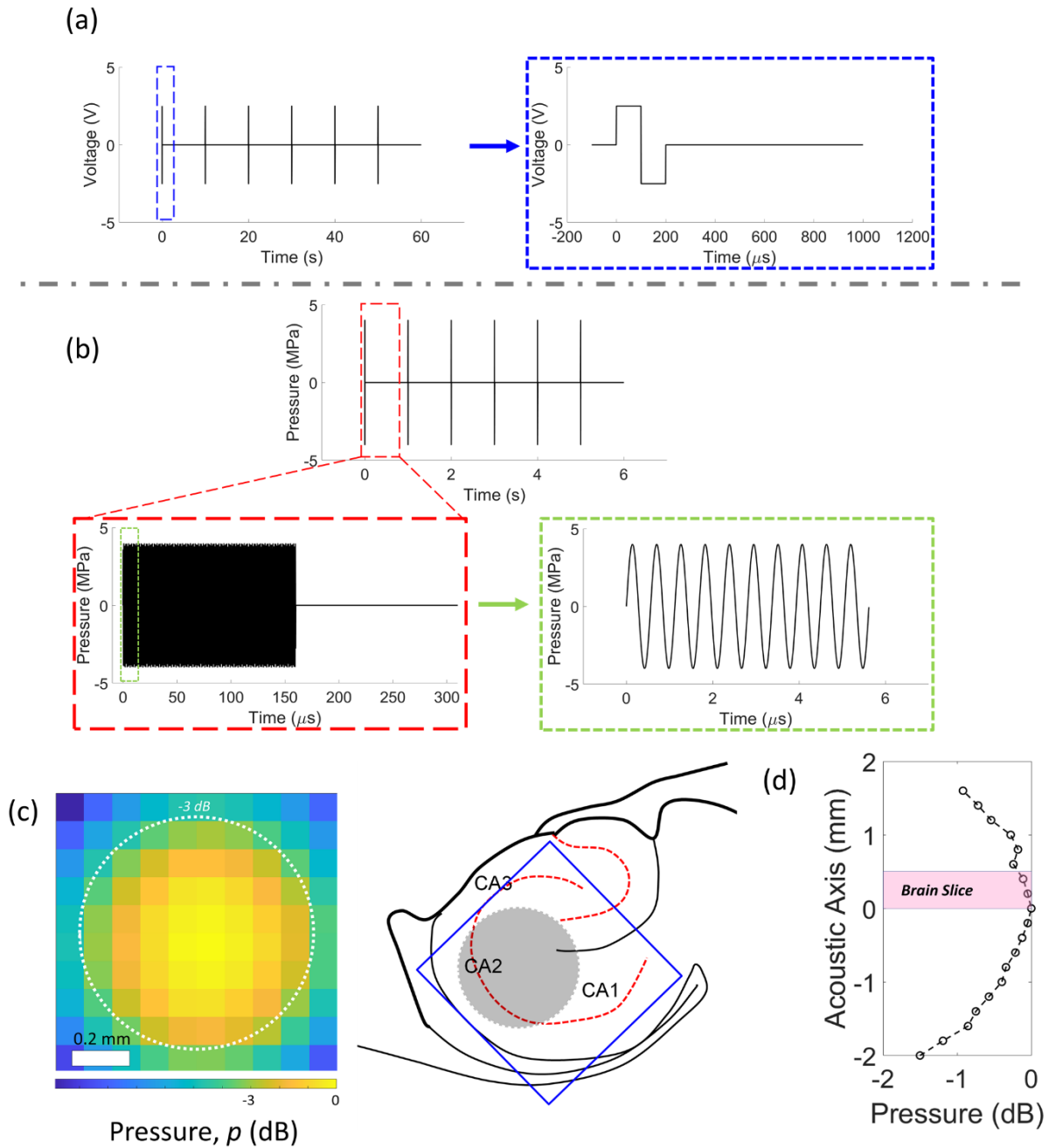

**Figure 3** – *(a) Example of temporal sequence utilized for electrical stimulation of hippocampal structures (5 Vpeak-peak biphasic pulse: 100 µs negative and 100 µs positive applied every 10 seconds) (b) Example of temporal sequence utilized for FUS stimulation of hippocampal structures (160 µs FUS pulses applied every second at peak pressure of 4 MPa). (c) Left: FUS focal region at 1.78 MHz frequency and resolution of stimulation. Relative pressure (dB) map (hydrophone scanning measurements) and -3 dB focal zone. Right: Relative size of the focal zone compared to the hippocampal structures and MEA grid. The focal zone was preferentially positioned to target CA3. (d) Relative pressure (dB) measured along the acoustic axis in the focal area. Thickness of the brain slice represented in pink.*

The experimental setup, in its current form, is subject to constructive and destructive interference patterns produced by reflections of the ultrasound waves at the aCSF-MEA interface underneath the hippocampal brain slice, and the aCSF-air interface. Indeed, the reflection coefficient between the neural tissue and the MEA glass was estimated to be 0.78. To estimate the effect of these reflections on the acoustic pressures applied on the brain slice, the experimental setup was modeled using a linear elastic model in COMSOL Multiphysics modeling software (version 5.4, COMSOL AB, Stockholm, Sweden). Material properties, parameters and dimensions were obtained from our own measurements, manufacturers' data, published data, public databases such as the nondestructive testing (NDT) resource ([www.nde-ed.org](www.nde-ed.org)) or the COMSOL database. The grid size was set to one fifteenth of the exposure wavelength to ensure proper spatial resolution as recommended by previous studies (Hensel et al., 2011). The simulation was run for the 160 µs duration of a single FUS pulse for a timestep set at 5 ns. A 1.78 MHz pressure sinewave of 284 cycles (pulse duration of 160 µs) and 176 kPa RMS pressure at the transducer surface was applied to the model. The pressure field simulated for the experimental setup was compared to that simulated in a free water field (**Figure 4a**). The spatial-average RMS pressure $p_{sar}$ ± standard deviation was calculated along the thickness of modelled brain slice over circular surfaces of 0.78 mm² (radius: 0.5 mm, slice thickness sampled at 100 µm) while including and excluding the MEA chip (**Figure 4b**).

Potential temperature elevations produced by 1.78 MHz FUS stimulation, applied at increasing FUS RMS pressures of 3, 4.1, 5.3, 6.3 MPa, 8.8 MPa and pulse durations of 200 µs, were estimated by using an in-house developed linear model software (ABLASIM) (Chavrier et al., 2000). FUS pressure of 8.8 MPa was included in the simulations to account for a potential doubling of the reflected acoustic power at the MEA glass surface. The material properties used for modelling the experimental setup are included in **Table 1**. The FUS transducer spatial sampling was set to one fifth of the FUS exposure wavelength and the modelled experimental setup was sampled at 0.1 mm. Only the FUS transducer, the aCSF (considered as water), the brain slice and the glass MEA were included in these simulations as the ABS cone was expected to have a negligeable effect on the heating of the brain slice.

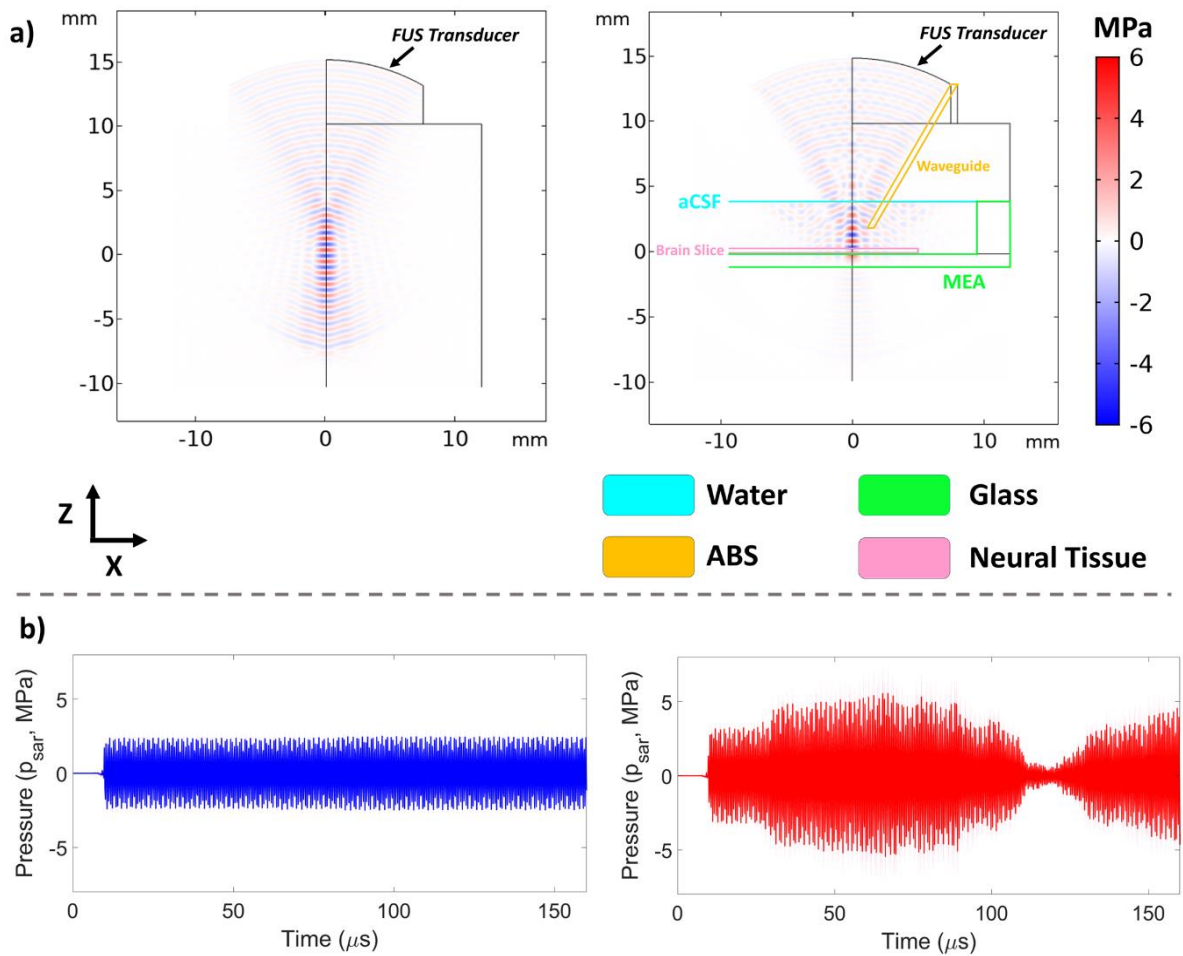

**Figure 4** – *(a) Left: 2-D pressure field in in a free water medium created using finite-element modelling with COMSOL Multiphysics. Right: 2-D pressure field in the experimental setup that includes the glass MEA chip (green), the ABS waveguide (orange), the brain tissue (pink), the aCSF modelled as water. (b) Spatial-average RMS pressure ($p_{sar}$) ± standard deviation calculated in the modelled brain tissue over a circular surface of 0.78 mm$^2$ (radius: 0.5 mm, at the equivalent surface of the brain slice: 400 µm from the MEA) with the MEA chip (right figure) and in over the same surface in a free water field (left figure).*

The physical parameters for all materials used in the finite-modelling simulations described in **Figure 4** are included in **Table 1**.

**Table 1** – *Properties of materials used for finite-element modelling*

| Material | Water | Glass | ABS | Neural Tissue |
|---|---|---|---|---|
| **Density (kg/m³)** | 1000 | 2210 | 1050[a] | 1040[c] |
| **Sound Velocity (m/s)** | 1500 | 5570[a] | 2250[a] | 1552[c] |
| **Acoustic Impedance (MRayls)** | 1.5 | 12.3[a] | 2.4[a] | 1.6[c] |
| **Attenuation (dB/cm)** | - | 16.2[b] | - | 1.7[c] |
| **Thermal Conductivity (W/m/K)** | 0.627 | 1.4 | - | 0.51[d] |
| **Specific Heat (J/kg-K)** | 4188 | 792 | - | 3600[d] |

[a] NDT Resource Center

[b] (Mashinskii, 2008)

[c] (Azhari, 2010)

[d] The Foundation for Research on Information Technologies in Society (IT'IS) ([www.itis.swiss](www.itis.swiss))

As expected, the pressure delivered to the brain slice is affected by the reflections created at the MEA-water interface as observed by the visibly disturbed pressure signal simulated with the MEA chip (**Figure 4b** – right graph) as compared to a free water field (**Figure 4b** – left graph). At the surface level displayed in the figure, there appears to be significant constructive interference along the duration of the FUS pulse, which increases the pressure to which the surface is subjected to. The pressures ± standard deviation calculated in a 0.31 mm$^3$ volume of the brain slice while including the MEA chip were 2.6 MPa ± 1.7, 2.1 MPa ± 1.4, 0.8 MPa ± 0.6, 1.1 MPa ± 0.8 and 2.4 MPa ± 1.7 in 0.78 mm$^2$ surface planes located along the thickness of brain slice (along the Z axis) at 0, 100, 200, 300 and 400 µm away from the MEA, respectively. In contrast, the pressures ± standard deviation calculated in a free field of water were 1.5 MPa ± 0.8, 1.5 MPa ± 0.8, 1.5 MPa ± 0.7, 1.5 MPa ± 0.8 and 1.6 MPa ± 0.8 for the same surface planes located 0, 100, 200, 300 and 400 µm away from the MEA, respectively. These results highlight that constructive and destructive interference patterns are produced at different levels across the thickness of the brain slice. The thermal simulations performed with ABLASIM calculated maximum temperature elevations in the brain slice of 0.07, 0.15, 0.25, 0.34 and 0.69 °C for 200 µs long FUS pulses at pressures of 3, 4.1, 5.3, 6.3 MPa and 8.8 MPa, respectively. These results are consistent with temperature elevations (< 1°C) previously reported in experimental setups similar to ours (Rinaldi et al., 1991; Bachtold et al., 1998). In these studies, the authors did not find evidence of neural activity modulation resulting from FUS produced temperature elevations of this magnitude.

### Comparison Electrical vs FUS Stimulation

Causal electrophysiological responses generated by single FUS pulses were compared to those generated by single pulse electrical stimulation. Indeed, literature on focused ultrasound (FUS) neurostimulation provides evidence of neuronal activity modulated by trains of repetitive FUS pulses/bursts (Tyler et al., 2008; King et al., 2012; Menz et al., 2013). However, various techniques such as DBS, TMS and TCS utilize short duration, single energy pulses for stimulation of neuronal activity (Miniussi et al., 2012): a strategy replicated in this work with FUS. Both stimuli were applied on the same single hippocampal brain slice. FUS stimulation was delivered as 160 µs, 6.3 MPa pulses applied at PRF of 1 Hz. For studies of synaptic transmission and neural connectivity along Schaffer collaterals as stimulated by electric biphasic pulses, it is common practice to stimulate pyramidal neurons in CA3 and to measure the transmitted responses downstream in CA1 (Chever et al., 2016). As such, electrical stimulation was applied along the Schaffer collaterals to elicit a response downstream in CA1.

Stimulation intensity was increased up to the steady state response in CA1. Single-pulse electrical stimulations (n = 50, 5 V$_{peak-peak}$ biphasic pulse: 100 µs negative and 100 µs positive, PRF: 0.1 Hz) was applied to avoid synaptic plasticity (**Figure 3a**). The response signals that were used for comparing both FUS and electrical stimulation were those recorded by the same electrode located on the Schaffer collaterals near CA1. The neurostimulation success rate (NSR), defined as the number of FUS or electrical pulses producing an LFP response to their total number of applied pulses, was calculated for both stimulating modalities after verifying successful stimulation of the brain slice.

### Inhibition of FUS-Induced Responses with TTX

In a sperate experiment, we ran a trial with Tetrodotoxin (TTX) to inhibit synaptic responses produced by FUS stimulation thus serving as a control through partial or full block of sodium (Na$^{2+}$)-channels. In the experiment, we applied FUS pulses at a PRF of 4 Hz and 6.3 MPa pressure at the focus. To obtain a baseline with high FUS-response repeatability (neurostimulation success rate of 100%), the pulse duration was increased to 3.5 ms for this experiment only. The first part of the experiment consisted in establishing a base LFP response stimulated by FUS pulses for approximately 90 seconds. At t = 90 s, we stopped the FUS stimulation and started perfusing the MEA chip with aCSF complemented with 0.5 µM TTX for the following 600 s. At t = 690 s, we restarted FUS stimulation using the exact parameters used to establish the baseline and maintained it for the final 200 s. The amplitude of the fEPSPs generated by FUS stimulation at baseline and after perfusion with TTX was compared.

### Temporal Variability and Spatial Distribution of FUS-Stimulated LFPs

Describing the temporal variability encountered in FUS-generated LFPs is important for the spatio-temporal analysis of FUS-stimulated electrical activity in the hippocampal brain slice. Therefore, in this section, we aimed at showcasing differences in the LFPs that can be generated as a result of stimulating FUS pulses. More precisely, we sought to describe, through various examples, the temporal variability that can be encountered between LFP responses to consecutive FUS pulses, and between LFP responses recorded by adjacent electrodes. Furthermore, we performed a control experiment aimed at confirming that the recorded LFPs were generated by the applied FUS pulses and not by the electromagnetic interference observed during each FUS pulse. The control experiment consisted in analyzing the signals recorded on a brain slice when FUS pulses (6.3 MPa, 160 µs, PRF = 1 Hz) were applied while centered on the MEA matrix (and thus the hippocampal structure), and when FUS pulses were applied away from brain slice. We expected to record LFPs only in the case where the FUS pulses were being directly delivered onto the hippocampal network.

Additionally, the spatial distribution of the FUS-generated LFPs across the MEA matrix was further explored in this section. FUS stimulation was applied on a single hippocampal brain slice as 6.3 MPa, 160 µs pulses delivered with a PRF of 1 Hz. The spatial distribution of the voltage recorded by all 60 MEA electrodes was reconstructed, which allowed for ultrafast 2D mapping of the signals recorded in the brain slice. This allowed for visualization of the voltage recorded by all electrodes in the MEA at different stages of the FUS-stimulated LFP.

## Comparison of FUS-Stimulated LFP Characteristics Among Different Brain Slices

The characteristics of the FUS-induced LFPs were compared among five different hippocampal brain slices from 4 different animals. In all five case, the FUS focal spot was centered on the brain slice while trying to aim at CA3 region of the Schaffer collaterals. Electrode recording the maximal fEPSP amplitude was identified and its characteristics were compared to those recorded in CA1 and CA3 and providing the maximal amplitude in these subregions. The FUS treatment applied to the different slices consisted of 160 – 200 µs pulses with 5.8 – 8 MPa pressures at the focus and PRF of 0.1 – 1 Hz. The pressures utilized to treat every slice were selected as that necessary to stimulate visible and repeatable responses across the MEA matrix. The LFPs recorded in electrodes located in CA3 and CA1 were compared to those exhibiting the strongest response in the MEA matrix as determined by their fEPSP amplitudes. The characteristics of the LFPs that were measured to make this comparison were the fEPSP amplitudes, delays, durations and slopes.

## Spatial Distribution of FUS-Stimulated LFP Characteristics Through Consecutive FUS Pulses

The spatial distribution of several fEPSP characteristics resulting from several consecutive FUS pulses was further studied to identify patterns of activation. In this trial, several consecutive 160 µs, 8 MPa FUS pulses were delivered on a single hippocampal brain slice at a PRF of 1 Hz. The characteristics of the LFPs stimulated by several consecutive FUS pulses, and that were thus measured as part of this analysis, were the fEPSP artifact amplitudes, fEPSP amplitudes, fEPSP delays and fEPSP slopes. These measured characteristics were then mapped to their respective recording electrodes in the MEA and 2D maps were generated for the responses triggered by the consecutive FUS pulses.

## Variability of FUS-Stimulated LFP Characteristics as a Function of Applied FUS Pressure

In a separate set of experiments, the evolution of responses created by FUS pulses of increasing pressures was examined. This trial was performed on a single slice and the LFPs produced by increasing FUS RMS pressures of 3, 4.1, 5.3 and 6.3 MPa at the focus were recorded by electrodes located in the vicinity of CA1 and CA3 regions of the hippocampus. In all cases, FUS pulses were applied for durations of 160 µs and at PRF of 1 Hz. The position of the FUS transducer remained unchanged throughout the duration of the experiment and as the FUS pressure was progressively increased. The characteristics of the LFPs stimulated by the different FUS pressures and that were thus measured for comparative purposes were the fEPSP amplitudes, delays and slopes.

### Data Analysis

The software MC_DataTool (V 2.6.15 MultiChannel Systems GmbH, Reutlingen, Germany) was used to convert the recorded raw data into ASCII files for further import and analysis using MATLAB (R2020a, MathWorks, Natick, Massachusetts, USA). Code was written for temporal concatenation of the different files generated by MC_DataTool for one specific treatment, as well as for automatic measurements of different parameters characteristic of electrical activity generated by FUS and electrical pulses.

A typical local field potential (LFP) generated by a single FUS pulse and recorded by a single electrode is displayed in **Figure 5**. In general, FUS-stimulated responses are initially characterized by an abrupt negative deflection lasting exactly the duration of the FUS pulse and is therefore identified as the FUS artifact. After the artifact, FUS-stimulated responses appear as field fEPSPs preceded by FVs. Our analysis code written in MATLAB detected the peaks of fEPSPs and FVs of the LFPs generated by FUS stimulation. The delays of FV and fEPSPs were calculated as the time difference between the start of stimulation artifacts and their respective peaks. The duration of the LFPs was calculated as the time necessary for the stimulated response to return to baseline. The slope of the fEPSP was calculated between points exhibiting voltages 10 to 25% superior to the absolute voltage at the base of the fEPSP and those exhibiting voltages 10 to 25% inferior to the absolute fEPSP peak (Lein et al., 2011). Given the noise level of the MEA recording system that was calculated and reported earlier in the methods section, we only included signals exhibiting peak fEPSPs higher than 50 μV.

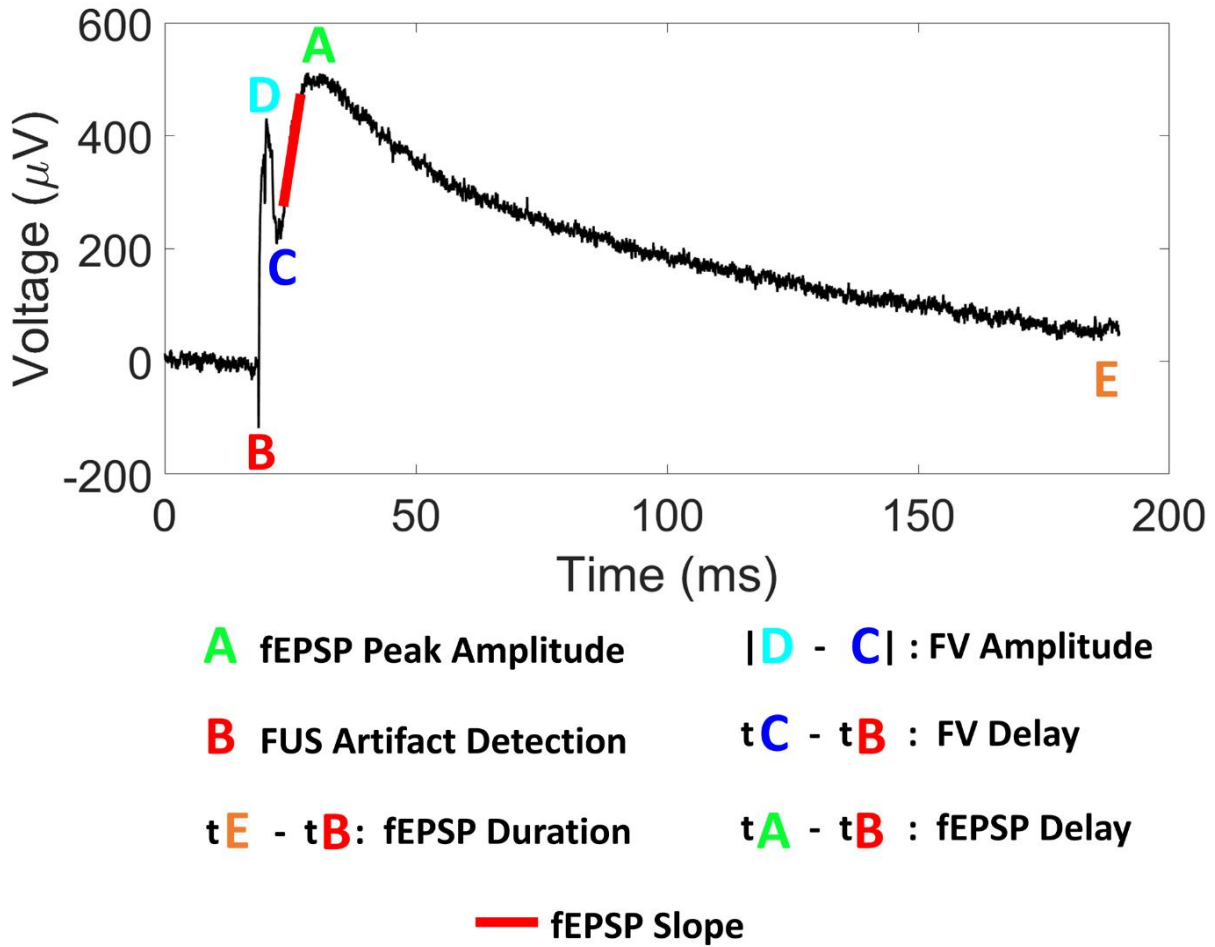

**Figure 5** - *Characteristic response generated by a single FUS pulse along with visual indicators of the signal parameters measured for spatio-temporal analysis of FUS-induced neural activity.*

Graph Pad Prism 8.3.1 Program (GraphPad Software, Inc., San Diego, CA, USA) was used for plotting and statistical analysis of collected data. Statistical analyses between FUS and electrically generated responses were performed using non-parametric Mann-Whitney tests. Statistical analyses comparing measurements on multiple hippocampal brain slice samples, as well as measurements from different FUS pressures levels applied to a single brain slice were performed using non-parametric Kruskal-Wallis tests. Statistical significance was established at $p < 0.05$.

### Division of Hippocampal Brain Slices Among Studies

For description of the different results obtained in this study, n is designated as the number of stimulated LFP responses that were evaluated in a particular study while m corresponds to the number of slices included in the analysis. The experiments in this work consisted of using 8 different brain slices (A, B, C, D, E, F, G, H) from 5 different animals (i, ii, iii, iv, v). The results presented in **Figure 6a** were generated using LFP responses as stimulated by FUS stimuli (n = 50) on slice G (m = 1) and electrical stimuli (n = 50) on slice H (m = 1). Comparison between electrically stimulated responses (n = 50) and

FUS-stimulated responses (6.3 MPa, 160 µs, PRF: 1 Hz, n = 25) were performed on slice C from animal ii (m = 1, **Figure 6b** through **d**). Experiments aimed at studying the effects of TTX on stimulation by FUS pulses (6.3 MPa, 5.5 ms, PRF: 4 Hz, n = 100: n = 50 before TTX and n = 50 after TTX) were performed on slice F (m = 1) from animal v (**Figure 7**). Spatial reconstruction of the voltage recorded by all 60 electrodes in response to a single FUS pulse (6.3 MPa, 160 µs, PRF: 0.1 Hz) to visualize the spatial distribution of the LFPs at different time points (**Figure 9**) was performed on slice A (m = 1, animal i). The comparison of FUS-stimulated LFP characteristics among different brain slices (**Figure 10**) were carried on slices A (6.3 MPa, 160 µs, PRF: 0.1 Hz, animal i), B (8 MPa, 160 µs, PRF: 1 Hz, animal ii), C (6.3 MPa, 160 µs, PRF: 1 Hz, animal ii), D (6.3 MPa, 200 µs, PRF: 1 Hz, animal ii) and E (5.8 MPa, 200 µs, PRF: 1 Hz, animal iii) for 10 LFPs each (n = 10 per slice, m = 5). The spatial distribution of FUS-induced LFP characteristics was analyzed over LFPs produced by 10 consecutive FUS pulses (n =10) from slice B (m = 1, animal ii, **Figure 11**). Finally, the study of the variability of FUS-stimulated LFP characteristics was done on slice C (m = 1, animal ii) (n ≥ 9 for each applied FUS pressure of 3, 4.1, 5.3 and 6.3 MPa, 160 µs, PRF: 1 Hz, **Figure 12**).

## Results

### Comparison Electrical vs FUS Stimulation

We first aimed at comparing the characteristics of the local field potentials (LFPs) generated by electrical stimulation to those generated by FUS stimulation. A direct comparison of the LFPs generated by both FUS and electrical pulses, at steady state, is displayed in **Figure 6**.

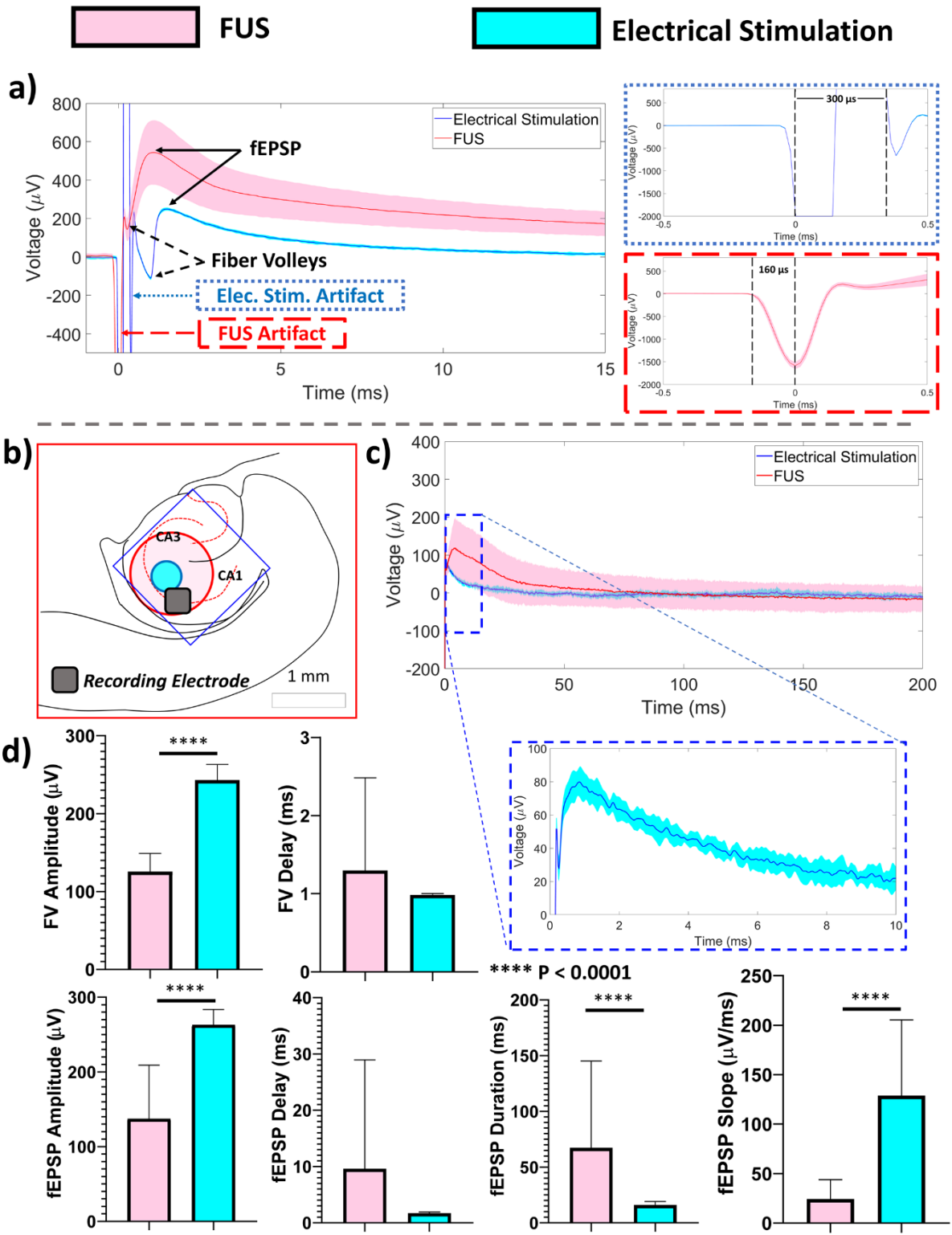

**Figure 6** – *Comparison of responses generated by FUS (mean (red curve) + standard deviation (pink filling)) and electrical stimulation (mean (blue curve) + standard deviation (cyan filling)).* **(a)** *Left: Graph comparing fEPSPs generated by FUS and electrical stimulation along with identification of principal components of the signal: stimulation artifacts, fiber volleys (FV), and fEPSPs. Right: Zoom over the electrical (top graph with blue, dotted outline) and FUS artifacts (bottom graph with red, dashed outline)* **(b)** *Localization of stimulating FUS (pink circle)*

*and electrical pulses (blue circle) and the recording electrode (gray circle) with respect to the brain slice anatomy. (c) Superposition of averaged fEPSPs generated by FUS stimulation (mean (red curve) + standard deviation (pink filling), n = 25 pulses) and electrical stimulation (mean (blue curve) + standard deviation (cyan filling), n = 50 pulses) along with a zoom on the response generated by electrical stimulation (blue dashed outline). (d) Statistical comparison of the FV and fEPSP characteristics in the responses generated by FUS and electrical stimulation: FV amplitude and delay; fEPSP amplitude, delay, duration and slope.*

The first step was qualitatively describing the LFPs stimulated by both FUS and electrical pulses. **Figure 6a** shows a superposition of responses generated by single FUS pulses and those generated by single electrical pulses. The responses to both electrical and FUS stimuli shown in this graph were selected because of their similar characteristics thus allowing the description of shared components such as FVs and fEPSPs. Exposure to both types of stimuli resulted in abrupt deflections that corresponded to their respective stimulation artifacts. In the case of the FUS artifact, the time between the start of the artifact and the deflection towards an eventual return to baseline (peak of the artifact) perfectly matches the duration of the FUS pulse (160 µs). At the end of the stimulation pulses, both signals experienced overshoots back to baseline followed by slow signals corresponding to fEPSPs. In these two cases, fEPSPs were preceded by FVs. However, despite displaying similar trends, the characteristics of both signals present significant differences.

To quantitatively compare the characteristics of LFPs produced by FUS (6.3 MPa, 160 µs, PRF: 1 Hz) and electrical pulses (5 Vpeak-peak biphasic pulse: 100 µs negative and 100 µs positive, PRF: 0.1 Hz), we evaluated the responses stimulated by both stimuli on the same hippocampal brain slice and recorded by the same individual electrode located on the Schaffer collaterals (**Figure 6b**). The mean responses to several FUS (red curve) and electrical (blue curve) pulses are displayed in **Figure 6c** where the lighter colors represent the variability of both signals as measured by the standard deviation. Statistical analysis of the response characteristics from both stimuli confirmed significant differences in the signals produced by FUS and electrical stimulation pulses (**Figure 6d**). Indeed, the only parameter that showed no statistical significance was the delay of the FV (electrical stimulation: $0.99 \pm 0.02$ ms, n = 50; FUS: $1.50 \pm 1.4$ ms, n = 25; $p > 0.05$). fEPSPs generated by FUS stimulation, in this particular case, appear to have higher delays ($20.2 \pm 28$ ms, n = 25, $p > 0.0001$) and durations ($128 \pm 96$ ms, n = 25, $p > 0.0001$) than fEPSPs generated by electrical stimulation (fEPSP delay: $1.7 \pm 0.2$ ms, fEPSP duration: $16 \pm 3$ ms, n = 50). However, electrically-stimulated fEPSPs exhibited steeper slopes ($372 \pm 88$ µV/ms, n = 50, $p > 0.0001$) and higher fEPSP and FV Amplitudes (fEPSP Amplitude: $262 \pm 21$ µV, FV Amplitude: $243 \pm 20$ µV, n = 50, $p > 0.0001$) than those generated by FUS stimulation (fEPSP slope: $29.6 \pm 31$ µV/ms, fEPSP Amplitude: $167 \pm 102$ µV, FV Amplitude: $130 \pm 35$ µV, n = 25). In addition to exhibiting different response characteristics to those stimulated by electrical stimulation, FUS stimulation had a NSR of 51% while electrical stimulation had a NSR of 100%. These results particularly underline the great variability found in LFPs generated by FUS stimulation as evidenced by the large standard deviations of their respective measurements.

### Inhibition of FUS-Induced Responses with TTX

The results of the control experiment involving the addition of $Na^+$ channel blocker TTX are displayed in **Figure 7.** The left column of the figure provides the results of FUS-induced LFPs at baseline before perfusing the hippocampal brain slice with TTX-supplement aCSF. The stability of base response over several consecutive FUS pulses was maintained as evidenced by the low standard deviation of the FUS-induced responses before addition of TTX. The right column presents the results of stimulating the brain slice with the same FUS parameters used to establish a base LFP response after continuously bathing the slice with TTX-supplement aCSF for 12 min. While the average fEPSP amplitude established at baseline was measured to be $124.3 \pm 16.8$ μV ($n = 50$, $p < 0.0001$), the effects of the TTX are highlighted in the results shown on the right column of **Figure 7** as it is evident that the FUS-stimulated fEPSP has completely disappeared after blockage of the $Na^+$ channels.

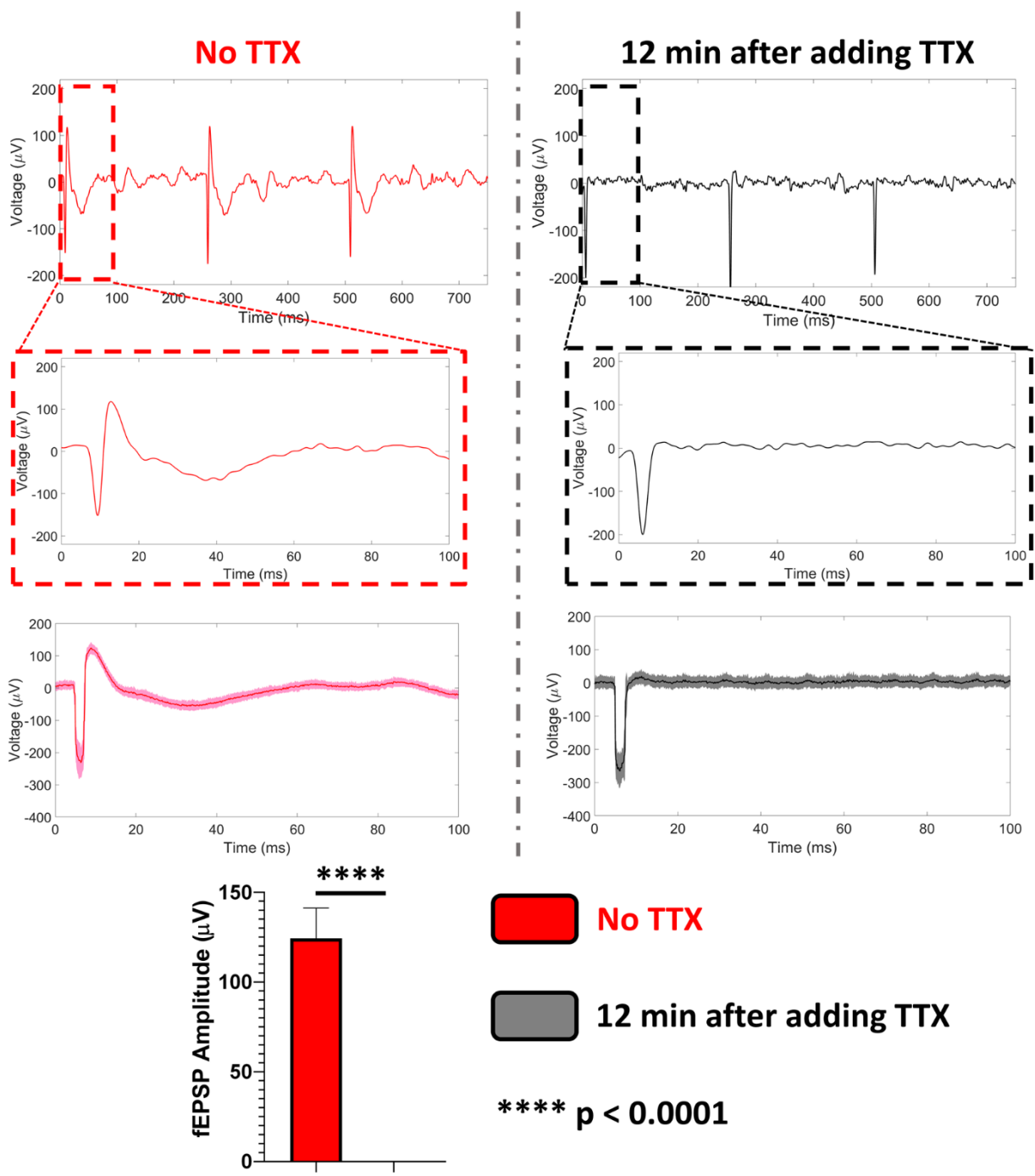

**Figure 7** – *Effects of TTX on FUS-induced LFPs in the hippocampal formation. Graphs are separated into two columns displaying the recorded signals by the same electrode within the MEA matrix before and 12 min after the addition of TTX. The graphs on the left column (red) correspond to the LFPs generated by FUS stimulation before perfusing the hippocampal slice with TTX while the graphs on the right column display the signals recorded after multiple FUS pulses 12 min after perfusing the slice with TTX. Graphs on the top row show the signals recorded during ten consecutive FUS pulses before (red) and after addition of TTX (black). The graphs in the middle provide a zoom of the signal recorded after a single FUS pulse in both respective cases. The graphs in the bottom row provide the mean ± standard deviation for consecutive signals recorded after several FUS pulses before (red, n = 50) and 12 min after addition of TTX-supplemented aCSF (black, n = 50). Histogram at the bottom of the figure provides a statistical comparison between the mean fEPSP amplitude ± standard deviation of FUS-elicited*

*responses before addition of TTX (n = 50) with those generated after 12 min of perfusion with the Na⁺ channel blocker (n = 50).*

## Temporal Variability and Spatial Distribution of FUS-Stimulated LFPs

Neuronal activities in the form of LFPs, generated as a result of stimulation by FUS pulses and exhibiting varying characteristics, are showcased in **Figure 8**. This figure shows the temporal signals recorded by single electrodes and displays responses to 10 consecutive FUS pulses applied at a PRF of 1 Hz. After the FUS-induced artifact, FUS-stimulated responses generally appear as fEPSPs exhibiting varying amplitudes and different polarities, depending on electrode location compared to neuronal layer. Indeed, **Figure 8a** displays a case where all fEPSPs were positive, **Figure 8b** displays a case where evoked fEPSPs were negative. Variations in stimulated fEPSPs are reflected in the graphs displaying the mean signals ± their respective standard deviations (right-most graphs, n = 10). **Figure 8c** shows the signals recorded by the entire electrode array in response to a single FUS pulse with particular emphasis on two adjacent electrodes exhibiting fEPSPs with different polarities. The statistical analyses performed throughout this study were thus applied on series of fEPSPs of similar polarities for the purpose of consistency. Furthermore, only FUS pulses that generated responses were included in these analyses. Indeed, contrary to electrical stimulation providing neurostimulation NSRs of 100%, we found that FUS stimulation exhibited NSRs ranging from 51% to 100%.

The results of the control experiment aimed at demonstrating that the observed LFP responses were caused by a direct interaction between the ultrasound beam and the neural structure, and not by the FUS artifacts, are presented in **Figure 8d.** When the FUS focal spot was not properly centered on the recording site, we see that the recorded signal includes only the artifact that is ever-present during FUS activation. However, when the FUS focal spot was carefully placed over the hippocampus and the recording electrode matrix, we observe the appearance of LFPs as a result of FUS pulses. Furthermore, it is worth noting that no LFP signal was recorded by any of the MEA matrix electrodes in the case where the FUS focus was not properly placed on the hippocampal structure (data not shown).

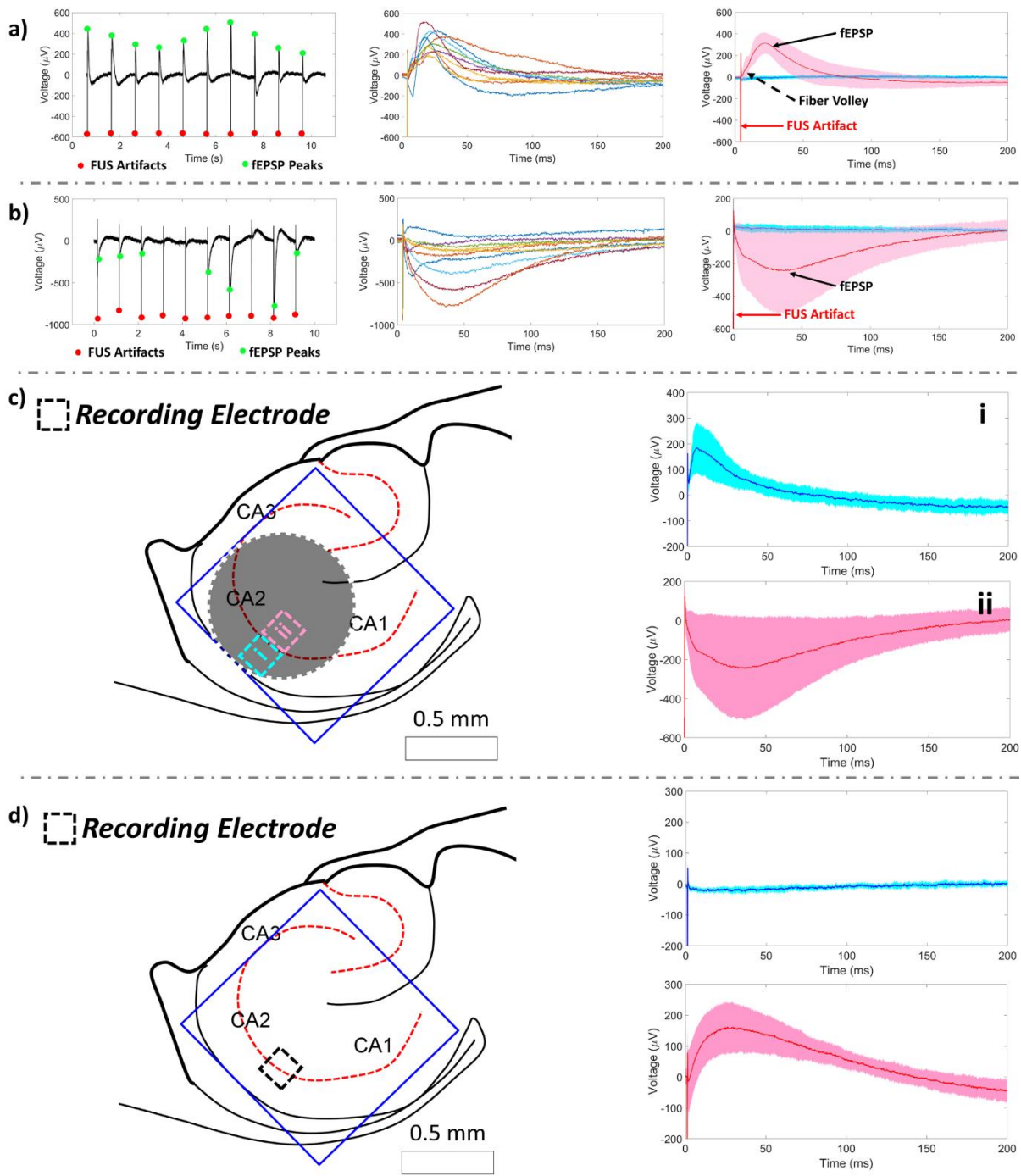

**Figure 8** – *Description of fEPSPs generated after 10 consecutive FUS pulses. **(a)** FUS-stimulated responses exhibiting only positive fEPSPs (n = 10), (b) negative fEPSPs (n = 10). Left-most graphs in **(a), (b)** correspond to continuous recordings of electrical signals during 10 consecutive FUS pulses along with indicators of the FUS artifacts (red circles) and resulting fEPSP peaks (green circles). Middle graphs superimpose the signals recorded immediately following the ten FUS pulses. Right-most graphs showcase the mean (red curve) ± standard deviation (pink fill) of the 10 fEPSPs displayed in their respective middle graphs. For comparative purposes, averaged FUS-induced fEPSPs are superimposed to averaged signals of a neighboring electrode that did not record any responses (blue curve). **(c)** Right: Mean responses ± standard deviations recorded by two adjacent electrodes on the matrix to the left, showcasing positive fEPSPs (blue curve) and negative fEPSPs (red curve). Left: Location of*

*the two recording electrodes within the hippocampal structure, MEA matrix and FUS focal spot for the two signals showcased on the right. **(d)** Right: Mean responses ± standard deviations recorded by a single electrode of the same slice while the FUS treatment was centered on the hippocampal structure (red curve) and when the FUS treatment was centered away from the hippocampal structure (blue curve). Left: Location of the recording electrode within the hippocampal structure and selected focal spot locations for treatment (pink circle) and control (blue circle).*

To visualize the response distribution across the MEA matrix resulting from a single FUS pulse (6.3 MPa, 160 µs), the spatial distribution of the voltage recorded by all 60 MEA electrodes was reconstructed, which allowed for ultrafast 2D mapping of the signals recorded in the brain slice (**Figure 9**). The map is overlaid on a picture of the treated brain slice placed over the MEA electrode matrix and taken with the camera included in the MEA system (**Figure 2a**). Six time-points are represented in the figure: a) Pre-stimulation baseline at t = 0 ms, b) FUS artifact at t = 20.2 ms, c) FV peak at 26.2 ms, d) fEPSP peak at t = 42.6 ms, e) progressive return to baseline at t = 100 ms and f) final return to baseline at t = 188 ms.

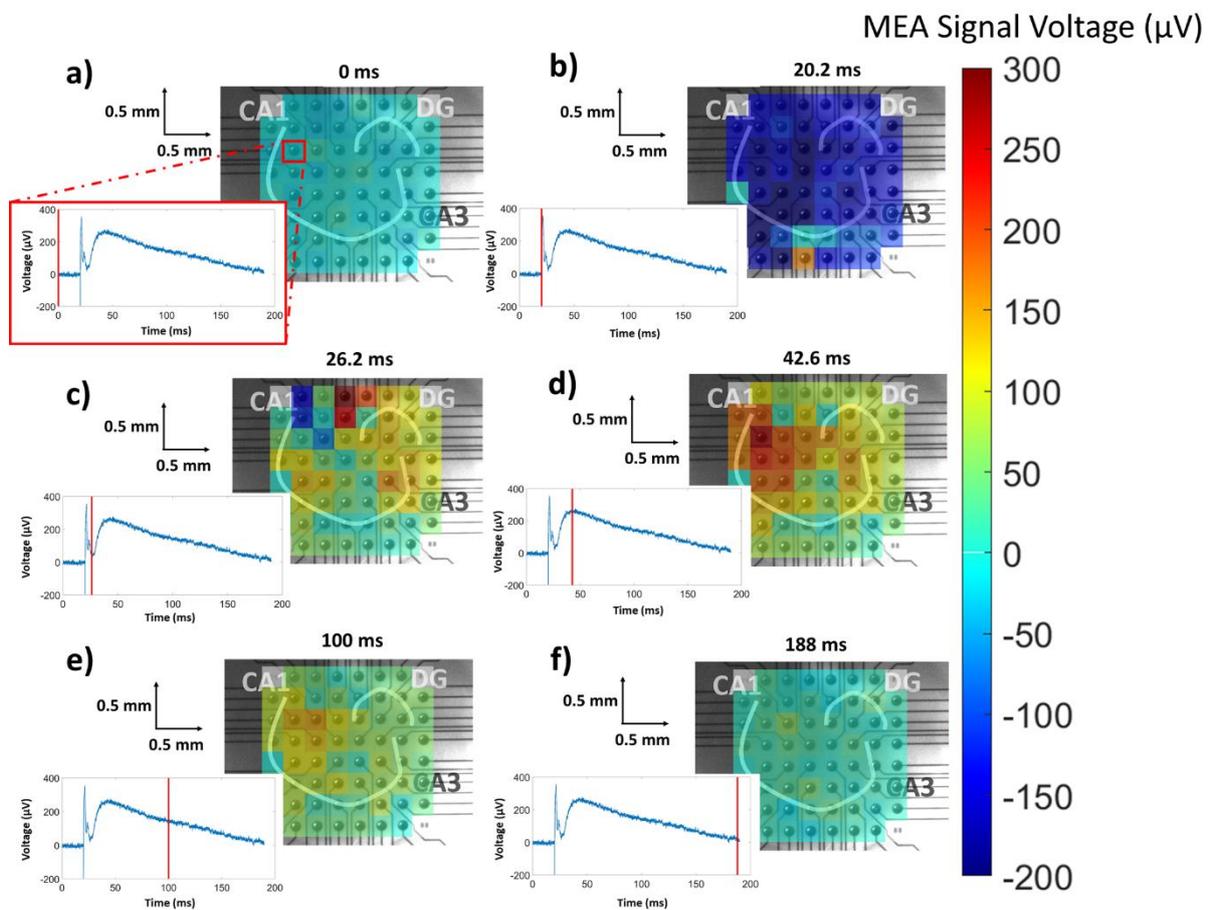

**Figure 9** - *Reconstruction of the FUS-evoked signals for all 60 electrodes within the MEA chip at six different time points **(a)** 0 ms (before FUS stimulation), **(b)** 20.2 ms (FUS artifact – negative deflection), **(c)** 26.2 ms (Negative*

*deflection corresponding to a Fiber Volley), **(d)** 42.6 ms (Positive deflection attributed to a neural response in the form of an fEPSP), **(e)** 100 ms (Slow, progressive return to baseline) and **(f)** 188 ms (Complete return to baseline), along with a characteristic electrical signal recorded in electrode 23 of the matrix near CA1 in the hippocampal brain slice (time point of the subfigure represented by vertical red line in 2D plot).*

At pre-stimulation state (**Figure 9a**), all electrodes recorded baseline potentials close to zero. At t = 20.2 ms (**Figure 9b**), a FUS pulse was applied which resulted in a fast reduction of the recorded potential in all electrodes corresponding to the FUS artifact and represented by colder colors across the entire MEA matrix. The FUS artifact was followed by a fast FV (**Figure 9c**), related to the action potential propagated through CA3 axons, Schaffer's collaterals, and a subsequent fEPSP, related to the downstream synchronized excitatory synaptic input in CA1 neurons (**Figure 9d**: peak amplitudes ≈600 µV at t = 42.6 ms). Though many electrodes within the MEA matrix appear to exhibit a response to the FUS pulse, there appears to be one region in particular displaying stronger responses as highlighted by the warmer colors in the upper-left quadrant of the MEA matrix. When placing this strongly responsive region within the anatomy of the hippocampal brain slice, it appears to be located towards the CA1 region of the hippocampus. Several tens of ms later, the electric potential recorded by all electrodes slowly returned to pre-stimulation baseline levels (**Figure 9e and f**).

## Comparison of FUS-Stimulated LFP Characteristics Among Different Brain Slices

Given the spatial and temporal variability in FUS-induced responses highlighted in **Figure 8** and **Figure 9**, specific spatial differences between fEPSPs triggered by stimulation centered on Schaffer's collaterals were explored. Electrode recording the maximal fEPSP amplitude was identified and its characteristics were compared to those recorded in CA1 and CA3 and providing the maximal amplitude in these subregions. **Figure 10** a) through e) provide averaged responses obtained in five different hippocampal brain slices (m = 5, slices A through E, respectively) treated with single 160 – 200 µs FUS pulses with 5.8 – 8 MPa RMS pressures at the focus. The pressures utilized to treat every slice were selected as that necessary to stimulate visible and repeatable responses across the MEA matrix. fEPSPs generated in CA3 (green curves) and CA1 (red curves) regions, along with the response exhibiting the highest fEPSP amplitude (blue curves) are displayed for all five brain slices. The location of the recording electrodes in each brain slice is indicated in schematics of the hippocampal brain slice.

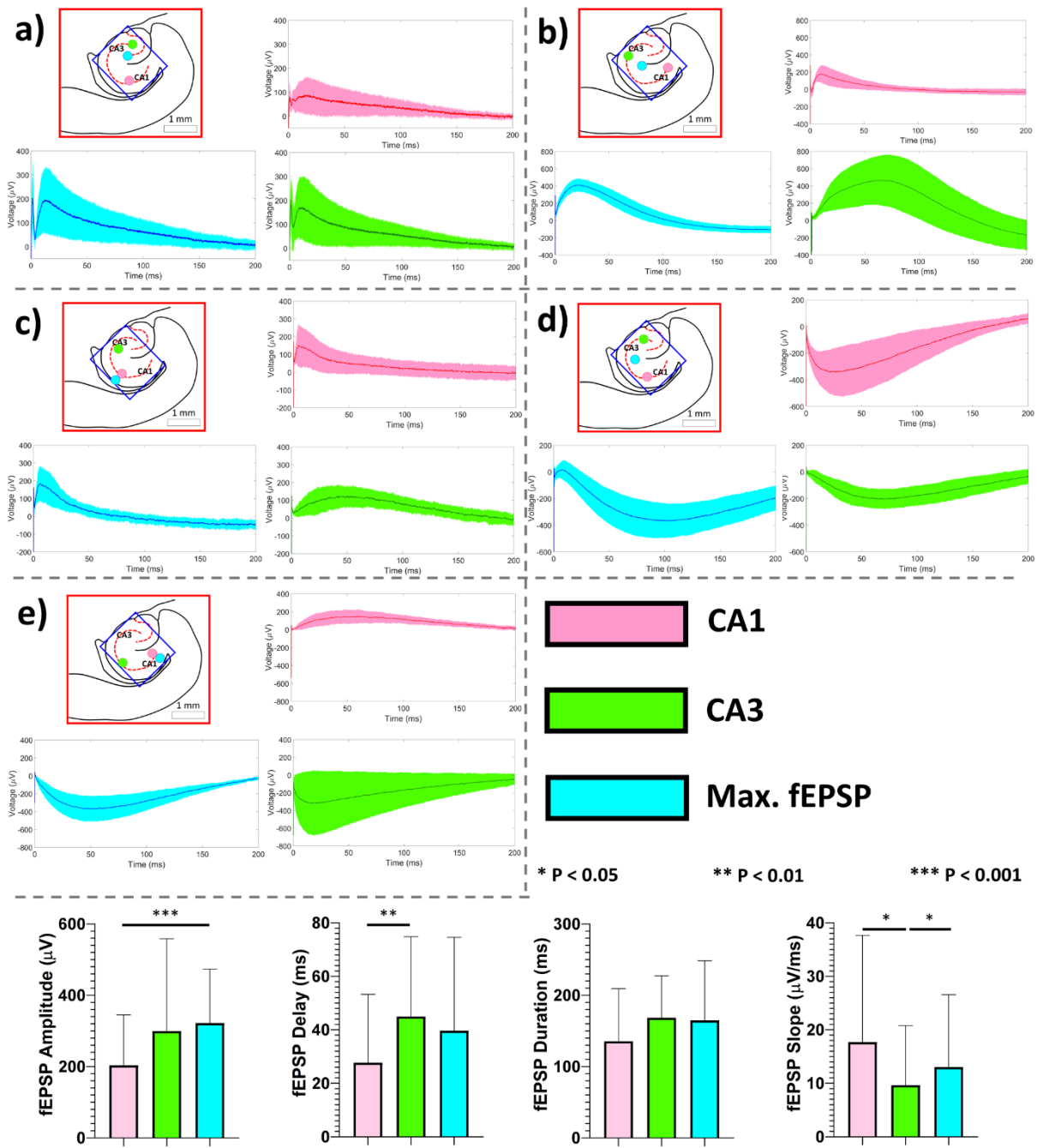

**Figure 10** – *Averaged characteristic signals (n = 10) recorded in CA1 (red), CA3 (green) and the maximum recorded signal (blue) in 5 different brain slices (m = 5, a through e). Each hippocampus schematic contains an estimated localization of the MEA matrix (blue square) and the recording electrodes in CA1 (pink), CA3 (green) and the maximum fEPSP amplitude. Statistical analyses of different fEPSP parameters are included in the bottom of the figure.*

The results in **Figure 10** show that FUS stimulation is capable of generating LFPs in both CA3 and CA1 and to activate bundles of axons in Schaffer's collaterals. Furthermore, the responses exhibiting the highest responses do not necessarily correspond to regions containing major neural pathways such as the Schaffer collaterals or the Mossy fibers, but involves a propagation pattern since most fEPSP peaked several tens of ms after stimulation and FV indicating action potential propagation through Schaffer's collaterals. When measuring the characteristics of these different responses, fEPSPs measured in CA3 tend were found to exhibit higher amplitudes (299 ± 258 µV, n × m = 50), delays (44.9 ± 29.9 ms, n × m = 50) and durations (168 ± 58.9 ms, n × m = 50) than fEPSPs measured in CA1 (fEPSP amplitude: 204 ± 141 µV, fEPSP delay: 27.7 ± 25.5 ms, fEPSP duration: 135 ± 73.7 ms, n × m = 50), while FV recorded in CA1 were the first sampled event (delay: 2.08 ± 1.45 ms, n × m = 30). Among these measurements, however, only the differences fEPSP delay were found to be statistically significant ($p > 0.01$). Nonetheless, the fEPSP slope in CA1 (177 ± 19.9 µV/ms, n × m = 50, $p > 0.05$) was measured to be statistically higher than that in responses measured in CA3 (9.66 ± 11.1 µV/ms, n × m = 50), suggesting a more synchronized activation.

The results described in **Figure 10**, taken together with the size of the FUS focal spot with respect to the MEA electrode matrix show in **Figure 3**, suggest that FUS treatment may have an effect on the majority of the hippocampal circuitry thus making FUS a global stimulation method as opposed to highly-localized electrical stimulations, especially in such conditions with a wide FUS stimulation of the hippocampal slice. Such feature of FUS stimulation may explain why we are able to record LFPs in different locations across the hippocampus. However, the dynamic of activation of hippocampal subregions validates that the whole slice was not activated directly by the stimulation and that neuronal activities may propagate within the slice. As a result, the spatial distribution of several parameters typical of FUS-induced fEPSPs was reconstructed and studied in an effort to describe the spatial characteristics of FUS-stimulated responses across the entire hippocampal brain slice.

## Spatial Distribution of FUS-Stimulated LFP Characteristics Through Consecutive FUS Pulses

The spatial distribution of several fEPSP characteristics to five consecutive FUS pulses is displayed in **Figure 11**. On the far-left column (red column) of **Figure 11a**, it can be seen that the FUS stimulation artifact is strongly present across the MEA matrix in all five pulses. Given that the time of flight of the FUS wave front is only 10 µs, it is possible that there is a positive relationship between the amplitude of the FUS artifact and the region of the MEA matrix exhibiting the strongest fEPSP responses. The second column from the left in **Figure 11a** (green column) shows the distribution of the measured fEPSP amplitude in response to five FUS pulses. fEPSPs of different amplitudes are detected by a great majority of the MEA electrodes. However, the central electrodes appear to detect the highest responses after every FUS pulse as highlighted by the warmer coloring (≈ 500 µV). When taking into consideration the anatomy and neural connectivity of the hippocampal brain slice, this would correspond to CA1 dendritic arborization. The second column from the right of **Figure 11a** (blue column) displays the delay between the stimulation artifact and the fEPSP peak across the MEA matrix. Higher fEPSP delays, as represented by warmer colors, were mostly measured away from the region exhibiting the highest fEPSP amplitudes, mainly towards the Schaffer collaterals. The distribution of the fEPSP slope across the matrix appears to follow a similar distribution to that of fEPSP amplitudes in that steeper slopes were measured in central regions located between CA1 and CA3 (right-most map in **Figure 11a**).

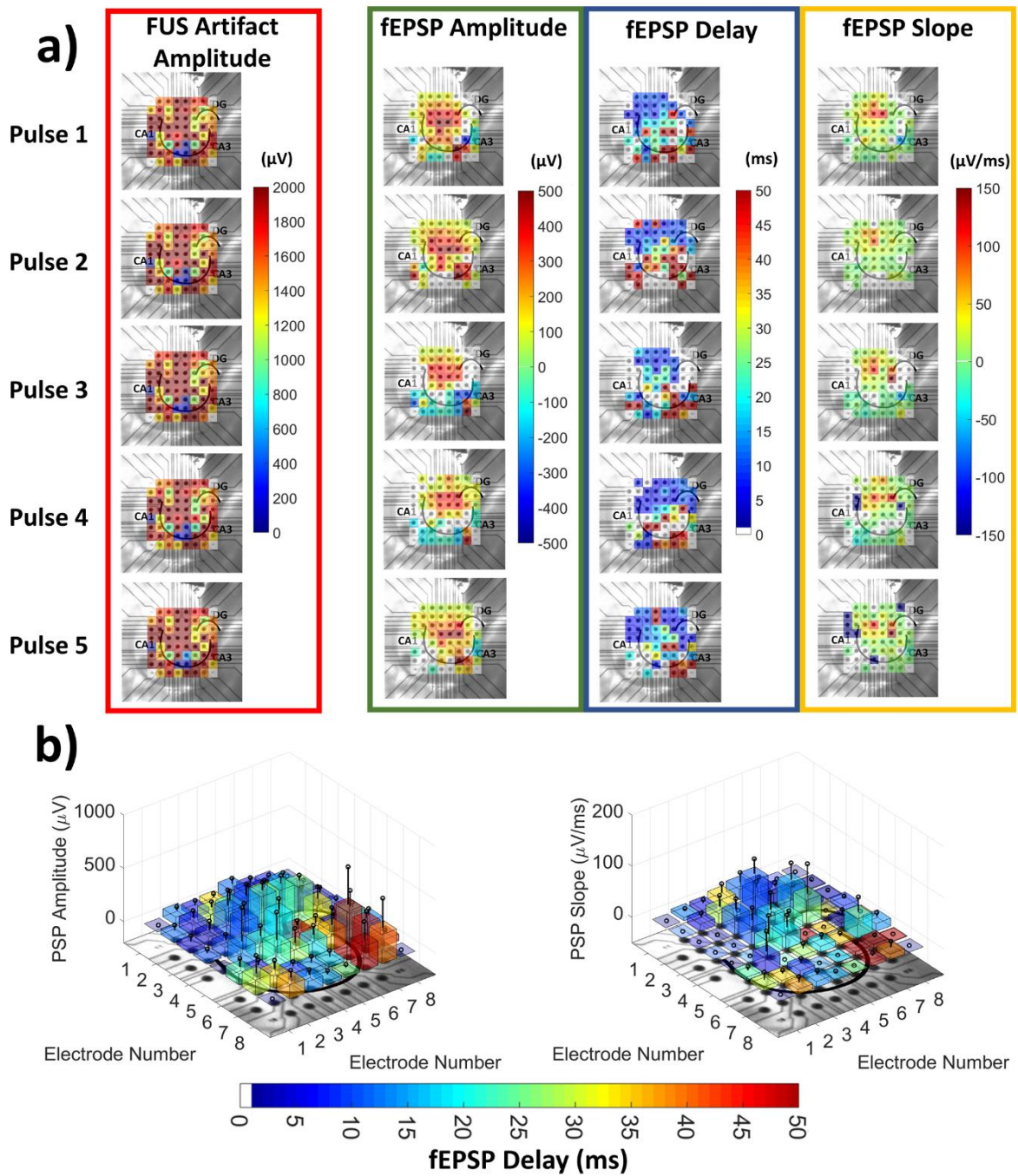

**Figure 11** – *(a) Spatial distribution of FUS stimulation artifact, fEPSP amplitude, fEPSP delay and fEPSP slope in response to five different, consecutive FUS pulses. (b) 3D spatial representation of fEPSP amplitude (left, n = 10, mean ± std) and fEPSP slope (right, n = 10, mean ± std) averaged over 10 consecutive FUS pulses. The delay of each fEPSP is represented by the color heat map.*

**Figure 11b** contains graphs providing the mean ± standard deviation (n = 10) of fEPSP amplitude (left graph) and slope (right graph) across the MEA matrix while displaying the average fEPSP delay as a color heat map. In this figure, it can be observed that higher fEPSP amplitudes and slopes in CA1 dendritic area and higher fEPSP delays towards the Schaffer collaterals in CA3 were consistent over ten consecutive FUS pulses.

## Variability of FUS-Stimulated LFP Characteristics as a Function of Applied FUS Pressure

**Figure 12** shows the mean ± standard deviation of the LFPs produced by FUS RMS pressures of 3, 4.1, 5.3 and 6.3 MPa at the focus in an electrode located near CA1 and CA3 of the hippocampus.

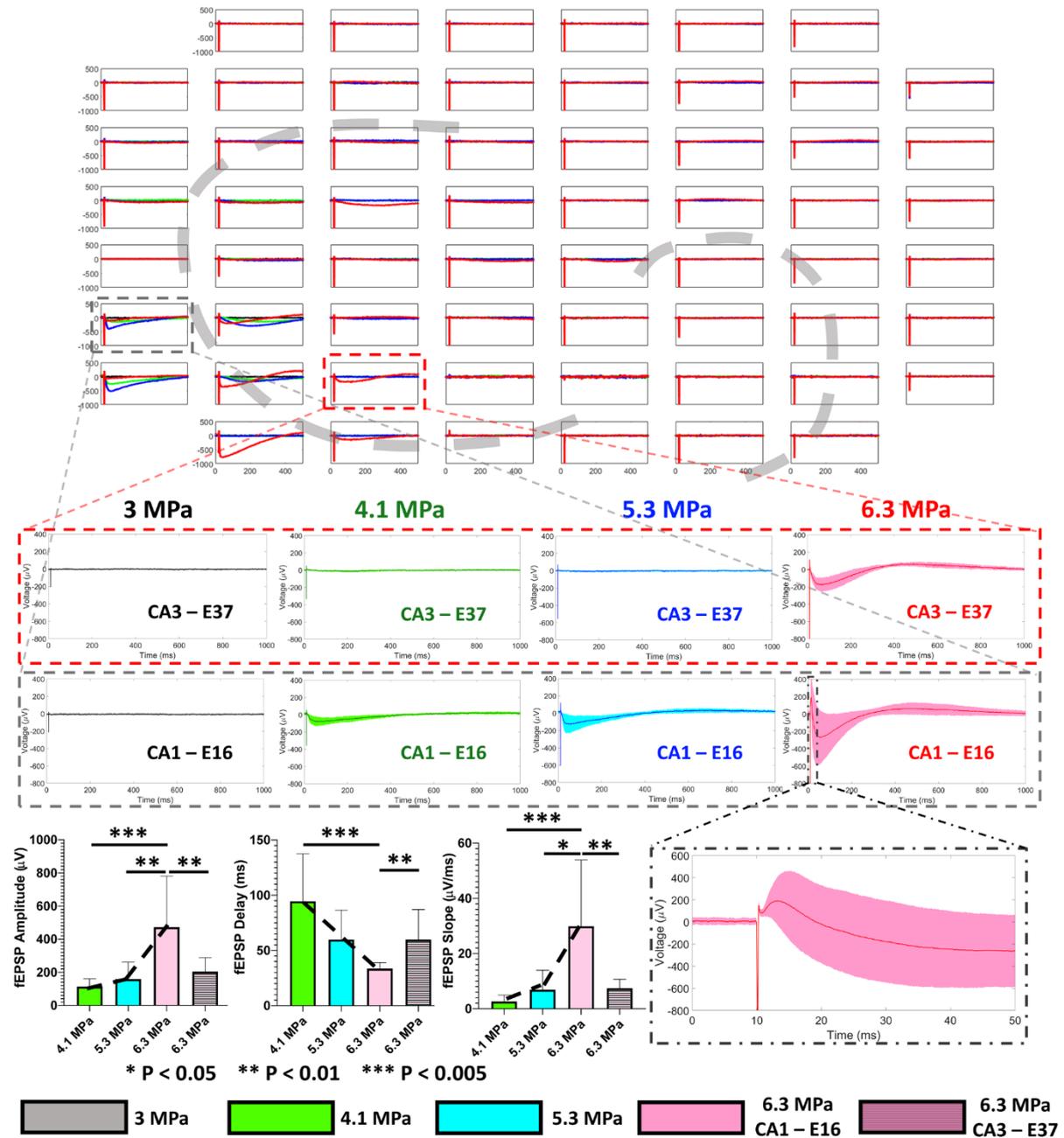

**Figure 12** – *Signals recorded in CA1 after 10 consecutive FUS pulses applied at pressures of 3, 4.1, 5.3 and 6.3 MPa at the focus (black, green, blue and red curves, respectively). An electrode located in the vicinity of CA1 was selected for comparing the responses generated by FUS pulses of varying pressures (Electrode 16: Gray, dashed enclosing). Additionally, an electrode located on the Schaffer collaterals further towards CA3 (Electrode 37: Red, dashed enclosing) was selected for comparison of its FUS-induced response to that recorded by electrode 16 in CA1. A zoom was included to the response generated in CA1 by 6.3 MPa pulses to better observe the emerge of fast responses at this level of FUS pressure. Statistics of the fEPSP amplitude, delay and slope (mean ± standard deviation) are included for FUS pulses that were capable of generating LFPs (4.1 MPa: Green, 5.3 MPa: Blue, 6.3 MPa in CA1: Pink, 6.3 MPa in CA3: Pink + horizontal lines).*

By applying increasing FUS pressures, increasingly stronger neuronal activity in the brain slice was generated in an electrode located in CA1 region of the Schaffer collaterals. Indeed, no activity appears to be recorded when applying 3 MPa FUS pulses. However, weak and slow responses begin to appear when increasing the pressure at the FUS focus to 4.1 MPa (fEPSP amplitude: 104 ± 36.4 µV, fEPSP delay: 86.9 ± 55.3 ms, fEPSP slope: 2.2 ± 1.3 µV/ms, n = 10). These responses further increased in amplitude and slope when elevating the FUS pressure to 5.3 MPa (fEPSP amplitude: 167 ± 103 µV, fEPSP slope: 7.5 ± 7.1 µV/ms, n = 11), while the fEPSP delay appears to have decreased (fEPSP delay: 55.1 ± 22.8 ms, n = 11), suggesting a stimulation intensity-dependent recruitment and synchronization of downstream neuronal populations. Finally, not only do we measure higher fEPSP amplitudes and slopes as we increase the FUS pressure to 6.3 MPa (fEPSP amplitude: 472 ± 306 µV, fEPSP slope: 2.4 ± 3.8 µV/ms, n = 9) along with shorter fEPSP delays (fEPSP delay: 33.5 ± 5.4 ms, n = 9), but we also observe the emergence of faster deflections at the beginning of such responses, which may correspond to detectable FVs since their mean amplitude and delay (FV amplitude: 77.18 ± 14 µV, FV delay: 0.88 ± 0.1 ms, n = 9) fall within values measured and displayed in **Figure 6**. The measured increase of fEPSP amplitudes, delays and slopes in responses generated by 6.3 MPa FUS pulses (n = 9), when compared to fEPSPs generated by 4.1 MPa FUS pulses (n = 10), was calculated to be statistically significant ($p < 0.005$). Similarly, fEPSP amplitudes and slopes were statistically higher in responses induced by 6.3 MPa pulses when compared to those generated by 5.3 MPa pulses ($p < 0.05$). The delay of the fEPSP was found to follow a negative linear relationship with respect to the FUS pressure at the focus ($R^2$ = 0.99), decreasing at a rate of approximately -35 ms/MPa. When looking at signals recorded by electrodes located along the Schaffer collaterals towards CA3, LFPs were only detected in response to FUS pulses with pressures of 6.3 MPa at the focus. The mean amplitude, delay, and slope of the fEPSPs produced in CA3 by consecutive FUS pulses of 6.3 MPa were 204 ± 83.9 µV, 59.9 ± 27.1 ms and 7.4 ± 3.2 µV/ms, respectively (n = 10). When compared to fEPSPs produced by FUS pulses in CA1, statistical significance was achieved for all three measurements ($p > 0.01$), suggesting that events recorded in CA3 were obtained for higher intensity of stimulation, with longer delays and with a less steep slope.

Interestingly, the amplitude of the FUS-induced artifact, measured on a single electrode, progressively increased with increasing FUS pressures as it was measured to be 214 ± 6 µV, 366 ± 15 µV, 611 ± 18 µV and 980 ± 53 µV for pressures of 3, 4.1, 5.3 and 6.3 MPa, respectively.

## Discussion

The results presented in this work demonstrate that neuronal synchronized signals in the form of fEPSPs can be stimulated from acute hippocampal brain slice as result of exposure to FUS. Electrophysiological responses resulting from FUS pulses were characterized by i) an initial sharp deflection to negative potentials lasting the exact duration of the pulse, thus corresponding to a FUS artifact created by the FUS exposure. The FUS artifact is followed by an immediate return to baseline and, in electrodes recording an electric response, ii) a fast negative deflection known as a FV, reflecting action potential propagation in the bundle of fibers of Schaffer's collateral (Sweatt, 2010) and iii) a slow fEPSP several milliseconds after the end of the FUS pulse, corresponding to a postsynaptic potential of neurons in the vicinity of the extracellular electrode (Sweatt, 2010). This response resembles that obtained by electrical stimulation as shown in **Figure 6**. Such recordings indicate that FUS allows production of a neuronal activity that secondarily propagates along neuronal structures, validating a neurostimulation process. The fEPSP generated by FUS stimulation also varied in polarity as we were capable of recording both positive and negative field potentials in electrodes located within the focal area produced by the FUS, suggesting more variable propagation and synchronization patterns. The differences in polarity arise from structural and functional factors in the hippocampus that combine to produce and shape different LFPs. The FUS-stimulated signals recorded throughout this study resemble those initially presented by Muratore et al. resulting from low-dose ultrasound application on neural networks and recorded with MEA systems (Muratore et al., 2009, 2012). Their study, however, did not describe the spatial characteristics and distribution of FUS induced LFPs nor did they describe the different signals arising from such treatment.

Recorded responses were shown to depend on neuronal activities since blocking voltage-gated sodium channels sustaining action potentials suppressed the response, as previously reported (Suarez-Castellanos et al., 2018). In our analysis, we found that the FUS-generated LFPs were entirely suppressed after 12 min of perfusion with TTX-supplemented aCSF as evidenced by the disappearance of the FUS-elicited fEPSP, thus suggesting that FUS-induced fEPSP are synaptically evoked. Furthermore, the stability of the response over several FUS pulses (n = 50) when establishing the baseline for these experiments indicates continuous functionality of the brain slice after repeated FUS pulses. An extended safety study aimed at evaluating the short- and long-term effects of FUS stimulation on brain slice function and viability will be explored in future studies.

With the studied parameters, FUS-generated LFPs exhibited a higher variability compared to LFPs generated by electrical stimulation as shown in **Figure 6**. The amplitude of potentials recorded by MEA electrodes is a function of the linear summation of surrounding neuronal sources and is inverse proportional to their distance (Obien et al., 2015). Therefore, the difference in amplitude may be the result of a higher number of neurons being simultaneously stimulated by a FUS pulse as compared to electrical stimulation. The focal spot size was measured to be approximately 0.8 mm in diameter, thus having surface area of 0.5 mm$^2$. Therefore, it can be safely assumed that a large area containing neurons and neural pathways located within the electrode matrix of the MEA chip (1.4 mm × 1.4 mm) will be exposed to the ultrasound beam. Hence, varying amplitudes and durations of fEPSPs and FVs elicited during FUS stimulations may correspond to varying number of neurons being activated by each stimulation event, or by variable pathways. Moreover, both excitatory and inhibitory circuits may be activated by FUS, contrarily to focal electrical stimulations of Schaffer's collaterals, which may sculpt variable neuronal activities within the hippocampus. Larger LFPs in specific regions may also be the result of a higher density of neural pathways responding to FUS stimulation. This can be seen within a single stimulation event across the entire MEA matrix, as electrodes located within certain regions display larger responses than those in surrounding electrodes as shown in **Figure 9**. In **Figure 11**, however, maximal fEPSP amplitudes and slopes in response to several FUS pulses were recorded towards the center of the MEA matrix which corresponds to an intermediate region of the hippocampus between CA1 and CA3, though away from the Schaffer collaterals, but which mostly samples CA1 dendritic signals. This raises the possibility that there is a positive relationship between regions exhibiting the highest fEPSP amplitudes and slopes and the region where the highest pressures within the FUS field are deposited. However, higher FUS frequencies, and hence smaller focal spots enabling better targeting capabilities, will be required to test this hypothesis.

Both FV delays and fEPSP delays in FUS-induced LFPs were shown to occur initially in CA1 but a clear fEPSP was sustainably recorded from CA3, as evidenced quantitatively in **Figure 10** and qualitatively in **Figure 11**. CA3 response was exhibiting larger amplitudes but was significantly delayed, suggesting they were generated by a secondary propagation. Further, they were underlined by significantly lower fEPSP slopes measured in CA3 as compared to CA1, suggesting a reduced synchronization process (**Figure 10**). Slower, larger events recorded in CA3 may also be due to a higher number of delayed currents from nearby neural structures congregating constructively around electrodes in CA3. It is also possible that neural bodies across the hippocampus may possess different sensitivities and excitabilities to FUS applied at specific parameters.

The current results do not describe how different FUS treatment parameters can modulate the characteristics of the generated LFPs. However, the effects of increasing the FUS pressure at the focus on the characteristics of the fEPSP response were demonstrated and presented in **Figure 12**. It was shown that LFPs of increasing amplitudes were produced as the focal RMS pressure of the FUS treatment was increased from 3 MPa to 6.3 MPa. While fifteen consecutive 3 MPa pulses failed to generate any detectable response, progressive elevation of the focal pressure produced LFPs exhibiting significantly higher fEPSP slopes and amplitudes. Furthermore, a fast response began to appear in response to 6.3 MPa pulses which may indicate the emergence of FVs. These results suggest the existence of an activation threshold for FUS pressures, an effect also observed in previous studies using other neural models (Menz et al., 2019; Vion-Bailly et al., 2019). However, Menz et al. were capable of stimulating responses at lower FUS pressures than those utilized in this study. Nonetheless, while the stimulation parameters commonly utilized in the field of FUS neurostimulation consist in the application several repetitive pulses, our stimulation protocol consisted of short single pulses which required higher FUS pressures. Indeed, different FUS-induced bioeffects may be responsible for triggering neural responses when applying repetitive or single FUS pulses. Furthermore, fEPSPs began appearing towards CA3 at higher FUS pressures which may be the result of linearly stimulating a higher number of neuronal sources in the vicinity of the Schaffer collaterals, but may also result of a propagation of CA1 activities beyond a threshold of recruitment. This observation is consistent with our measurements showing higher fEPSP amplitudes resulting from increasing FUS pressures. Thus, the emergence of FVs identified in CA1 at FUS pressures of 6.3 MPa may be caused by signals generated in CA3 and transmitted down the Schaffer collaterals. A parametric study of other FUS pulse characteristics (i.e. pulse duration, driving frequency, pulse repetition frequency) and their effect on generated LFPs is necessary for further characterization of this promising technology.

The described results confirm the use of a mixed MEA/FUS platform for spatio-temporal studies of neurostimulation/neuromodulation by FUS. Such platform can be used to characterize the electrochemical mechanisms regulating this phenomenon. The spatial and temporal contribution of other acting membrane proteins and ions could further be elucidated using this mixed setup. In particular, the role of calcium and respective channels in synaptic transmission could provide valuable insights into FUS signal generation and propagation across neuronal networks (Suarez Castellanos et al., 2016).

## Conclusion

In this study, it is demonstrated that MEA systems can be a useful tool for studying and characterizing the spatio-temporal biophysical and electrophysiological processes elicited by FUS stimulation in neuronal networks. These initial results show promise in the capabilities of MEA systems to not only elucidate the mechanisms involved in ultrasound neurostimulation and/or neuromodulation, but also provide initial insights into potential clinical applications.

## Acknowledgements

This project was supported by the French National Research Agency (ANR2016, N°ANR-16-TERC-0017), the Laboratory of Excellence (LabEx) DevWeCan and the Focused Ultrasound Foundation (LabTAU, Center of Excellence of the FUSF). The authors would like to acknowledge the work of Dr. Françoise Chavrier for simulating FUS-produced temperature elevations in our experimental setup.

## Author contributions

I. M. S. C. designed the study, conducted the experiments, carried out the simulations, processed the raw data, analyzed the results and wrote the manuscript. W.A.N. designed the study, conducted the experiments, provided critical feedback for the different steps of this study, analyzed the results and reviewed the manuscript. E. D. conducted the experiments and reviewed the manuscript. L. S. helped in the analysis of data the development of data analyzing tools. J. V. B. conducted the experiments, reviewed the manuscript. A. C. reviewed the manuscript. G. H. provided critical feedback in the different steps of this study and reviewed the manuscript. J. Y. C. reviewed the manuscript.

## Conflict of Interest

The authors declare no conflict of interest.